\documentclass[10pt,twocolumn,oneside,final]{IEEEtran}

\usepackage{cite}
\usepackage{graphicx}
\usepackage{amsmath}
\usepackage{times}
\usepackage{latexsym}
\usepackage{bm}
\usepackage{amssymb}
\usepackage[center]{caption2}
\usepackage{array}
\usepackage{setspace}
\usepackage{fancyhdr}
\usepackage{cite,graphicx,amsmath,amssymb}
\usepackage{citesort}
\usepackage{color}
\usepackage{multirow}
\usepackage{times}
\usepackage{amsmath}
\usepackage{amssymb}
\usepackage{amsfonts}
\usepackage{graphics}
\usepackage{amssymb}
\usepackage{booktabs}

\ifCLASSINFOpdf
\else
\fi

\makeatother

\begin{document}

\title{Downlink Performance of Pilot-Reused HetNet with Large-Scale Antenna Arrays}

\author{Yongyu~Dai,~\IEEEmembership{Student Member,~IEEE,}
        ~Xiaodai~Dong,~\IEEEmembership{Senior~Member,~IEEE,}
        and~Hai~Lin,~\IEEEmembership{Senior~Member,~IEEE}

\thanks{Y.~Dai and X.~Dong are with the Department of Electrical and Computer Engineering, University of Victoria, Victoria, V8P 5C2, Canada (Email: yongyu@uvic.ca, xdong@ece.uvic.ca). H.~Lin is with the Department of Electrical and Information Systems, Osaka Prefecture University, Sakai, Osaka, 599-8531, Japan (Email: lin@eis.osakafu-u.ac.jp).

        Part of this work has been presented at the IEEE Global Communications Conference (GLOBECOM), San Diego, CA, USA, Dec. 2015~\cite{Dai2015}.}}
\maketitle

%

\begin{abstract}

Considering a heterogeneous network (HetNet) where both macro base station (BS) and small cell (SC) nodes are equipped with massive antennas, this paper studies the performance for multiple-input multiple-output (MIMO) downlinks when the macro and small cells share the same spectrum and hence interfere with each other. Suppose that the large-scale antenna arrays at both macro BS and SC nodes employ maximum-ratio transmission (MRT) or zero-forcing transmission (ZFT) precoding, and transmit data streams to the served users simultaneously. A new pilot reuse pattern among SCs is proposed for channel estimation. Taking into account imperfect channel state information (CSI), capacity lower bounds for MRT and ZFT are derived, respectively, in closed-form expressions involving only statistical CSI. Then asymptotic analyses for massive arrays are presented under specific power scaling laws. Subsequently, two user scheduling algorithms, greedy scheduling algorithm and asymptotical scheduling algorithm (ASA), are proposed based on derived capacity lower bounds and asymptotic analyses, respectively. ASA is demonstrated to be a near optimal in the asymptotic regime and has low complexity. Finally, the derived closed-form expressions are verified to be accurate predictors of the system performance by Monte-Carlo simulations. Numerical results demonstrate the effectiveness of asymptotic analysis and proposed user scheduling schemes.

\end{abstract}


\IEEEpeerreviewmaketitle

\section{Introduction}\label{sec:introduction}

As a viable and cost-effective way to increase network capacity, heterogeneous networks (HetNets) that embed a large number of low-power nodes, called small cells (SCs), into an existing macro network has emerged with the aim to offload traffic from the macro cell (MC) to small cells~\cite{3gpp36.814,Lin2015,hetnet2011ltea,hetnet20113gpp,5G} in hot spots or to solve coverage holes in MC. Conventionally deploying more macro base stations (BSs) in already dense networks may be prohibitively expensive and result in severe inter-cell interference~\cite{hetnet20113gpp}. However, due to the large number of potentially interfering nodes in the network, mitigating both the inter-cell and intra-cell interference becomes a crucial issue facing HetNet. Interference control has been intensively studied and applied in HetNet~\cite{Gesbert2010,Zhu2011,Dai2013}, including the coordinated multi-point (CoMP) transmission \cite{Gesbert2010}. Although the CoMP transmission was shown to provide high spectral efficiency~\cite{CompHetNet2014} with the backhaul among the coordinated tiers enabling both user data and channel state information (CSI) exchange, the high signaling overhead results in practical implementation limitations.

Recently, multiple-input multiple-output (MIMO) transmission with large-scale antenna arrays at the BS has attracted substantial interest from both academia and industry. Using simple linear processing, such large-scale antenna arrays were proved to be able to substantially reduce the effects of the uncorrelated noise, small-scale fading and intracell interference~\cite{Marzetta2010,FF2013}. Then, the energy and spectral efficiency of very large multiuser MIMO uplink systems were investigated in \cite{Ngo2013}, which showed that the power radiated by each terminal could be made inversely proportional to either the number of BS antennas or at least its square-root, considering both perfect and imperfect CSI. In \cite{5G}, it was stated that the potential benefits have elevated large-scale MIMO to a central position as a promising technology for the next generation of wireless systems. In a HetNet setting, \cite{Hosseini2013,Adhikary1_2014} proposed to use large scale antenna arrays at the BS and limited antennas at the SCs\footnote{In this paper, we use SC to denote the SC node and BS to denote the macro cell node for simplicity.} due to their smaller form factor. As variable structures of antenna arrays, such as a cylindrical array, requires less space~\cite{Gao2011}, large-scale antenna arrays set at SCs becomes realizable.  Recently, NEC Corporation announced that it has developed a prototype of A4-sized massive-element Active Antenna System for 5G small cell base stations, and it proposed the use of a massive-element antenna in small cells for capacity enhancement~\cite{NECWhitePaper}.  Up to date, few papers in the literature have studied the effect of employing massive MIMO at SCs.  In \cite{Adhikary2_2014}, \emph{HetNet with large-scale antenna arrays} was investigated on downlink performance with interference coordination, and random matrix theory was used to simplify the analysis significantly. It was shown in~\cite{Dai2015} that using large-scale antenna arrays in SC reduces both intra-tier interference and the cross-tier interference from other nodes in the HetNet system, leading to higher spectral efficiency and better coverage, especially for hot zones.

This paper presents a comprehensive study of a two-tier network with large-scale antenna arrays set at both BS and SCs. In our preliminary literature~\cite{Dai2015}, maximum-ratio transmission (MRT) precoding was employed based on the estimated channels obtained from the orthogonal training scheme, and downlink capacity lower bounds for a user in the MC and for a user in an SC were derived in closed-form expressions. However, there are still many critical yet unsolved problems. This paper makes the following contributions to address the remaining issues.
\begin{enumerate}
\item It was stated in \cite{Marzetta2006} that pilot overhead is proportional to the number of user equipment (UE) for the conventional orthogonal training scheme, i.e., the system performance will degrade as the UE number grows due to heavy pilot overhead. In \cite{Marzetta2010,FF2013}, the pilot reuse (PR) technique is utilized among the macro cells to reduce the pilot overhead, while UEs within a cell use orthogonal pilots. \cite{Sun2015} studies pilot reuse in a dense small cell network. In a two-tier HetNet with multiple small cells, massive antenna arrays and large number of UEs, we propose to apply pilot reuse among the SCs in this paper, i.e., the same set of orthogonal pilots is reused among the small cells in one macro-cell. Thus the number of orthogonal pilots is smaller than the total UE number in the whole network.
\item We present for the first time the downlink capacity lower bounds of the large-scale HetNet system, where simple linear precoding such as MRT or zero-forcing transmission (ZFT) is employed at each node, followed by detailed asymptotic analysis.
\item The design of an efficient and practical user scheduler for the large-scale HetNet is an important and challenging problem, because the required CSI exchange becomes prohibitively complicated due to the large-scale antenna arrays and the large number of UEs in the MC and SCs. Based on the obtained capacity bounds and asymptotic analysis, a greedy scheduling algorithm (GSA) and an asymptotic scheduling algorithm (ASA) are proposed, respectively, where GSA requires only statistical CSI (SCSI) shared between BS and SCs, and ASA even removes the need for any CSI exchange among nodes.
\end{enumerate}

The rest of the paper is organized as follows. We briefly describe the system model for HetNet with large-scale antenna arrays in Section~\ref{sec:sys_model}. In Section~\ref{sec:Achiev_Rate}, lower bounds for the achievable rate are derived with both imperfect CSI based MRT and ZFT, followed by corresponding asymptotic analysis. Then, two user scheduling algorithms are developed in Section~\ref{sec:US}. Moreover, simulation results under different system configurations are given in Section~\ref{sec:NumericalResults} to demonstrate the effectiveness of both the derived rate expressions and the developed schemes. Finally, conclusions are drawn in Section~\ref{sec:conclusion}.

\emph{Notations:} For a matrix $\mathbf{X}$, we use ${\bf{X}}^T$, ${\bf{X}}^H$, ${\bf{X}}^*$ and ${\rm {Tr}}\{\bf{X}\}$ to denote the transpose, the Hermitian transpose, the conjugate, and the trace, respectively. ${\bf I}_N$ is an $N\times N$ identity matrix and ${\bf 0}_{M\times N}$ is an $M\times N$ zero matrix. Moreover, ${\rm {E}}[\cdot]$ denotes the expectation operator. The symbol $\left\|{\bf x}\right\|$ indicates the 2-norm of vector $\bf x$, and ${\rm {diag}}\{{\bf x}\}$ denotes a diagonal matrix with $\bf x$ being its diagonal entries. $\left|{x}\right|$ is the absolute value of $x$, while $\left|{U}\right|$ represents the number of elements in set $U$. Finally, the notation $\mathop  {\longrightarrow} \limits^{a.s.}$ means almost sure convergence, and ${\bf{x}} \sim \mathcal{CN}(0,{\bf D}_x)$ represents a circularly symmetric complex Gaussian vector $\bf x$ with zero mean and covariance matrix ${\bf D}_x$.

\section{System Model}\label{sec:sys_model}
\begin{figure}
   \centering
   \includegraphics[scale=0.3]{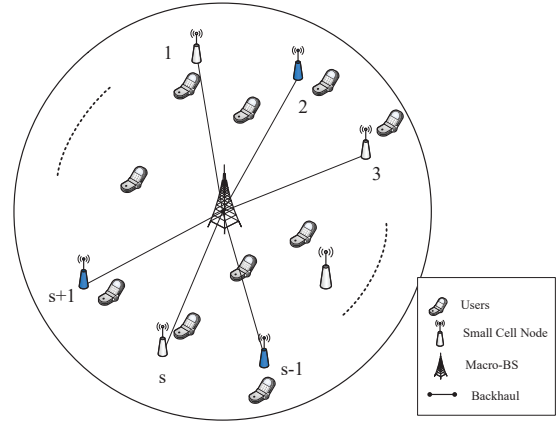}
   \caption{System model for HetNet with SCs deployment.}
   \label{Section3_1}
\end{figure}

Fig.~\ref{Section3_1} shows the considered two-tier network architecture with one cell consisting of one macro BS, which is overlaid with a dense tier of $S$ uniformly distributed SCs by sharing the same time-frequency resources. Assume that the BS and SCs are respectively equipped with large-scale arrays of $N_\text{BS}$ and $N_\text{SC}$ antennas, where $N_\text{BS} > N_\text{SC} \gg 1$, while each user has only one antenna due to the size or complexity constraint. Notably, uniform user distribution in the cell is focused here. Based on the biased user association~\cite{Lin2015}, the users served by the macro BS are designated to a macro UE (MUE) set, and those served by each SC are designated to a small cell UE (SUE) set. Furthermore, suppose that the macro BS serves $K$ MUEs simultaneously while each SC serves $L$ SUEs with $K\le N_\text{BS}$ and $L\le N_\text{SC}$. Denote the MUE and SUE sets as $U_{\text{M}}$ and $U_{\text{S}}^{(m)}$, respectively, then we have $K\le \left| U_{\text{M}} \right|$ and $L\le \left| U_{\text{S}}^{(m)} \right|$. The selected subsets of MUE and SUE after user scheduling are denoted by $I$ and $J_m$~($m\in\left\{1,\dots,S\right\}$), respectively.

For the channel matrices, they account for both small-scale fading and large-scale fading. Here, we assume that all the channels between the users and the nodes follow independent and identically distributed (i.i.d.) Rayleigh fading and time division duplex (TDD) is adopted with channel reciprocity satisfied. Denote the channel matrices from the BS and $n$th~($n\in\left\{1,\dots,S\right\}$) SC to the $K$ MUEs as ${{\bf{G}}_{{\rm{B - M}}}} = \left[ {{\bf{g}}_{{\rm{B - M}}}^{(1)}, \ldots ,{\bf{g}}_{{\rm{B - M}}}^{(K)}} \right]\in{\mathbb{C}^{N_\text{BS}\times K}}$ and ${\bf{G}}_{{\rm{S - M}}}^{(n)} = \left[ {{\bf{g}}_{{\rm{S - M}}}^{(n,1)}, \ldots ,{\bf{g}}_{{\rm{S - M}}}^{(n,K)}} \right]\in{\mathbb{C}^{N_\text{SC}\times K}}$, respectively, and use ${\bf{G}}_{{\rm{B - S}}}^{(m)} = \left[ {{\bf{g}}_{{\rm{B - S}}}^{(m,1)}, \ldots ,{\bf{g}}_{{\rm{B - S}}}^{(m,L)}} \right]\in{\mathbb{C}^{N_\text{BS}\times L}}$ and ${\bf{G}}_{{\rm{S - S}}}^{(n,m)} = \left[ {\bf{g}}_{{\rm{S - S}}}^{(n,m,1)}, \ldots ,\right.$ $\left.{\bf{g}}_{{\rm{S - S}}}^{(n,m,L)} \right]\in{\mathbb{C}^{N_\text{SC}\times L}}$ to represent the channel matrices from the BS and the $n$th SC to the $L$ SUEs in the $m$th SC, respectively. We have ${{\bf{G}}_{{\rm{B - M}}}} = {{\bf{H}}_{{\rm{B - M}}}}{\bf{D}}_{{\rm{B - M}}}^{{1 \mathord{\left/
 {\vphantom {1 2}} \right.
 \kern-\nulldelimiterspace} 2}}$, ${\bf{G}}_{{\rm{B - S}}}^{(m)} = {\bf{H}}_{{\rm{B - S}}}^{(m)}{\left( {{\bf{D}}_{{\rm{B - S}}}^{(m)}} \right)^{{1 \mathord{\left/
 {\vphantom {1 2}} \right.
 \kern-\nulldelimiterspace} 2}}}$, ${\bf{G}}_{{\rm{S - M}}}^{(n)} = {\bf{H}}_{{\rm{S - M}}}^{(n)}{\left( {{\bf{D}}_{{\rm{S - M}}}^{(n)}} \right)^{{1 \mathord{\left/
 {\vphantom {1 2}} \right.
 \kern-\nulldelimiterspace} 2}}}$ and ${\bf{G}}_{{\rm{S - S}}}^{(n,m)} = {\bf{H}}_{{\rm{S - S}}}^{(n,m)}{\left( {{\bf{D}}_{{\rm{S - S}}}^{(n,m)}} \right)^{{1 \mathord{\left/
 {\vphantom {1 2}} \right.
 \kern-\nulldelimiterspace} 2}}}$
where $n,~m\in\left\{1,\dots,S\right\}$, the first items ${{\bf{H}}_{{\rm{B - M}}}}\in{\mathbb{C}^{N_\text{BS}\times K}}$, ${\bf{H}}_{{\rm{B - S}}}^{(m)}\in{\mathbb{C}^{N_\text{BS}\times L}}$, ${\bf{H}}_{{\rm{S - M}}}^{(n)}\in{\mathbb{C}^{N_\text{SC}\times K}}$ and ${\bf{H}}_{{\rm{S - S}}}^{(n,m)}\in{\mathbb{C}^{N_\text{SC}\times L}}$ include the i.i.d. $\mathcal{CN}(0,1)$ small-scale fading coefficients, and the second items are the large-scale fading diagonal matrices given by ${{\bf{D}}_{{\rm{B - M}}}} = {\rm {diag}}\left\{ {\beta _{{\rm{B - M}}}^{(1)}, \dots ,\beta _{{\rm{B - M}}}^{(K)}} \right\}$, ${\bf{D}}_{{\rm{B - S}}}^{(m)} = {\rm {diag}}\left\{ {\beta _{{\rm{B - S}}}^{(m,1)},\dots ,\beta _{{\rm{B - S}}}^{(m,L)}} \right\}$, ${\bf{D}}_{{\rm{S - M}}}^{(n)} = {\rm {diag}}\left\{ {\beta _{{\rm{S - M}}}^{(n,1)}, \dots ,\beta _{{\rm{S - M}}}^{(n,K)}} \right\}$, and ${\bf{D}}_{{\rm{S - S}}}^{(n,m)}={\rm {diag}}\left\{ \beta _{{\rm{S - S}}}^{(n,m,1)}, \dots ,\right.$ $\left.\beta _{{\rm{S - S}}}^{(n,m,L)} \right\}$.

\subsection{Channel Estimation with Pilot Reuse}

Practically, the channel matrix from each node to its corresponding users, i.e., ${{\bf{G}}_{{\rm{B - M}}}}$ and ${\bf{G}}_{{\rm{S - S}}}^{(m,m)}$~($m\in\left\{1,\dots,S\right\}$), have to be estimated based on the uplink training. At the beginning of each coherence interval $T$, all users simultaneously transmit pilot sequences of length $\tau $ symbols. On account of the slight interferences between low-power SCs which are far away from each other, we present a pilot reuse pattern for small cells in a large-scale HetNet system.

First, we denote the reuse factor as $\gamma$, i.e., all SCs utilize $\gamma$ sets of $L$ pairwise orthogonal pilot sequences with a total of $SS=S/\gamma$ SCs sharing the same set. This requires $\tau \ge K+L\times \gamma$ to satisfy the orthogonality of the MC and SC pilot sets. Then, we group SCs into $\gamma$ sets according to the maximum relative distance criterion and SCs in one set use the same pilot sequences. Since all low-power nodes are modeled as uniformly distributed in a circle with BS at the center as shown in Fig.~\ref{Section3_1}, we can denote the $n$th~($n\in\left\{1,\cdots,\gamma\right\}$) SC set as ${\mathcal A}_n=\left\{n, n+\gamma,\cdots, n+(SS-1)\gamma\right\}$. Taking $S=8$ and reuse factor $\gamma=2$ for example, the SC sets are ${\mathcal A}_1=\left\{1, 3, 5, 7\right\}$ and ${\mathcal A}_2=\left\{2, 4, 6, 8\right\}$. The $4$ SCs in each set share one pilot set which includes $L$ pairwise orthogonal pilot sequences, and there are $2$ pilot sets for the total of $8$ SCs.

Then the training matrix received at the BS and the $m$th~($m\in{\mathcal A}_r$) SC can be written as
\begin{equation}\label{sec3:train_BS}
\begin{split}
{{\bf{Y}}_{{\rm{BS}}}} &= \sqrt {\tau {p_{\tau}}} \left({{\bf{G}}_{{\rm{B - M}}}}{{\bf{\Phi }}_{{\rm{MUE}}}} + \sum\limits_{t = 1}^{\gamma} {\sum\limits_{l \in {{\mathcal A}_t}} {{\bf{G}}_{{\rm{B - S}}}^{(l)}{\bf{\Phi }}_{{\rm{SUE}}}^{(t)}} }\right)  + {{\bf{N}}_{{\rm{BS}}}}\\
{\bf{Y}}_{{\rm{SC}}}^{(m)}& = \sqrt {\tau {p_{\tau}}} \left(\sum\limits_{t = 1}^{\gamma} {\sum\limits_{l \in {{\mathcal A}_t}} {{\bf{G}}_{{\rm{S - S}}}^{(m,l)}} {\bf{\Phi }}_{{\rm{SUE}}}^{(t)}}  + {\bf{G}}_{{\rm{S - M}}}^{\left( m \right)}{{\bf{\Phi }}_{{\rm{MUE}}}}\right) + {\bf{N}}_{{\rm{SC}}}^{(m)}
\end{split}
\end{equation}
respectively, where $p_{\tau}$ is the transmit power of each pilot symbol, ${{\bf{N}}_{{\rm{BS}}}}$ and ${\bf{N}}_{{\rm{SC}}}^{(m)}$ are the additive white Gaussian noise (AWGN) matrices with i.i.d. components following $\mathcal{CN}(0,\sigma_0^2)$, the training vectors transmitted by the $i$th~($i\in I$) MUE is denoted by the $i$th row of ${{\bf{\Phi }}_{{\rm{MUE}}}}\in{\mathbb{C}^{K\times \tau}}$, satisfying ${\bf{\Phi }}_{{\rm{MUE}}}{\bf{\Phi }}_{{\rm{MUE}}}^H={{\bf{I}}_K}$, while the training vector transmitted by the $j$th ($j\in J_m$) SUE of one SC in the $r$th set ${\mathcal A}_r$ is represented by the $j$th row of ${{\bf{\Phi }}_{{\rm{SUE}}}^{(r)}}\in{\mathbb{C}^{L\times \tau}}$, satisfying ${{\bf{\Phi }}_{{\rm{SUE}}}^{(r)}}\left({{\bf{\Phi }}_{{\rm{SUE}}}^{(r)}}\right)^H={{\bf{I}}_L}$. Moreover, since the rows of pilot sequence matrices are pairwise orthogonal, we have ${{\bf{\Phi }}_{{\rm{MUE}}}}\left({{\bf{\Phi }}_{{\rm{SUE}}}^{(r)}}\right)^H={{\bf{0}}_{K\times L}}$ and ${{\bf{\Phi }}_{{\rm{SUE}}}^{(r)}}\left({{\bf{\Phi }}_{{\rm{SUE}}}^{(t)}}\right)^H={{\bf{0}}_{L\times L}}$~($\forall r\ne t \in\{1,\dots,\gamma\}$).

In order to estimate ${{\bf{G}}_{{\rm{B - M}}}}$ and ${\bf{G}}_{{\rm{S - S}}}^{(m,m)}$~($m\in{\mathcal A}_r$), we employ the minimum mean-square-error (MMSE) estimation at each node~\cite{Kay1993}. The estimated channels are given by
\begin{equation}\label{sec3:estimated_CSI}
\begin{split}
{{{\hat { \bf  G}}}_{{\rm{B - M}}}} &= \frac{1}{{\sqrt {\tau {p_{\tau}}} }}{{\bf{Y}}_{{\rm{BS}}}}{\bf{\Phi }}_{{\rm{MUE}}}^H{{{\tilde{ \bf  D}}}_{{\rm{B - M}}}}\\
&= {{\bf{G}}_{{\rm{B - M}}}}{{{\tilde{ \bf  D}}}_{{\rm{B - M}}}} + \frac{1}{{\sqrt {\tau {p_{\tau}}} }}{{{\tilde{ \bf  N}}}_{{\rm{BS}}}}{{{\tilde{ \bf  D}}}_{{\rm{B - M}}}}\\
{\hat{\bf G}}_{{\rm{S - S}}}^{(m,m)}& = \frac{1}{{\sqrt {\tau {p_{\tau}}} }}{\bf{Y}}_{{\rm{SC}}}^{(m)}{\left( {{\bf{\Phi }}_{{\rm{SUE}}}^{(r)}} \right)^H}{\tilde{ \bf  D}}_{{\rm{S - S}}}^{(m,m)}\\
&= \sum\limits_{l \in {{\mathcal A}_r}} {{\bf{G}}_{{\rm{S - S}}}^{(m,l)}} {\bf{\tilde D}}_{{\rm{S - S}}}^{(m,m)} + \frac{1}{{\sqrt {\tau {p_\tau}} }}{\bf{\tilde N}}_{{\rm{SC}}}^{(m)}{\bf{\tilde D}}_{{\rm{S - S}}}^{(m,m)}
\end{split}
\end{equation}
where ${{{\tilde{ \bf D}}}_{{\rm{B - M}}}} \buildrel \Delta \over = {\left( {\frac{{{{\bf{D}}_{{\rm{B - M}}}^{-1}}}\sigma_0^2}{{\tau {p_{\tau}}}} + {{\bf{I}}_K}} \right)^{ - 1}}$, ${\tilde{ \bf D}}_{{\rm{S - S}}}^{(m,m)}\buildrel \Delta \over = {\left[ {\left( {\sum\limits_{l \ne m,l \in {{\mathcal A}_r}} {{\bf{D}}_{{\rm{S - S}}}^{(m,l)}}  + \frac{{\sigma _0^2}}{{\tau {p_{\tau}}}}{{\bf{I}}_L}} \right){{\left( {{\bf{D}}_{{\rm{S - S}}}^{(m,m)}} \right)}^{ - 1}} + {{\bf{I}}_L}} \right]^{ - 1}}$, ${{{\tilde{ \bf N}}}_{{\rm{BS}}}} \buildrel \Delta \over = {{\bf{N}}_{{\rm{BS}}}}{\bf{\Phi }}_{{\rm{MUE}}}^H$ and ${\tilde{ \bf  N}}_{{\rm{SC}}}^{(m)} \buildrel \Delta \over = {\bf{N}}_{{\rm{SC}}}^{(m)}{\left( {{\bf{\Phi }}_{{\rm{SUE}}}^{(r)}} \right)^H}$~($m\in{\mathcal A}_r$) are defined. Due to the property of ${{\bf{\Phi }}_{{\rm{MUE}}}}$ and ${\bf{\Phi }}_{{\rm{SUE}}}^{(r)}$, ${{{\tilde{ \bf  N}}}_{{\rm{BS}}}}$ and ${\tilde{ \bf  N}}_{{\rm{SC}}}^{(m)}$ are also composed of i.i.d. $\mathcal{CN}(0,\sigma_0^2)$ elements. Then, we have
\begin{equation}\label{sec3:estimated_CSI_exp}
\begin{split}
{{\bf{G}}_{{\rm{B - M}}}}& = {{{\hat { \bf G}}}_{{\rm{B - M}}}} + {{\bf{\Xi }}_{{\rm{B - M}}}},{\bf{G}}_{{\rm{S - S}}}^{(m,m)} = {\hat { \bf G}}_{{\rm{S - S}}}^{(m,m)} + {\bf{\Xi }}_{{\rm{S - S}}}^{(m,m)}
\end{split}
\end{equation}
where ${{\bf{\Xi }}_{{\rm{B - M}}}}$ and ${\bf{\Xi }}_{{\rm{S - S}}}^{(m,m)}$ denote the estimation error matrices which are independent of ${{{\hat { \bf G}}}_{{\rm{B - M}}}}$ and ${\hat { \bf G}}_{{\rm{S - S}}}^{(m,m)}$ from the property of MMSE channel estimation~\cite{Kay1993}. Hence, we have ${{{\hat { \bf G}}}_{{\rm{B - M}}}} \sim CN\left( {0,{{{\hat { \bf D}}}_{{\rm{B - M}}}}} \right)$ with ${{{\hat { \bf D}}}_{{\rm{B - M}}}} = {\rm {diag}}\left\{ {\hat \beta _{{\rm{B - M}}}^{(1)}, \ldots ,\hat \beta _{{\rm{B - M}}}^{(K)}} \right\}$, ${\hat { \bf G}}_{{\rm{S - S}}}^{(m,m)} \sim CN\left( {0,{\hat { \bf D}}_{{\rm{S - S}}}^{(m,m)}} \right)$ with ${\hat { \bf D}}_{{\rm{S - S}}}^{(m,m)}$ $= {\rm {diag}}\left\{ {\hat \beta _{{\rm{S - S}}}^{(m,m,1)}, \ldots ,\hat \beta _{{\rm{S - S}}}^{(m,m,L)}} \right\}$, ${{\bf{\Xi }}_{{\rm{B - M}}}} \sim CN\left( {0,{{\bf{D}}_{{\rm{B - M}}}} - {{{\hat { \bf D}}}_{{\rm{B - M}}}}} \right)$ with the $i$th column vector denoted by ${{\bf{\xi }}_{{\rm{B - M}}}^{(i)}}$, and ${\bf{\Xi }}_{{\rm{S - S}}}^{(m,m)} \sim CN\left( {0,{\bf{D}}_{{\rm{S - S}}}^{(m,m)} - {\hat { \bf D}}_{{\rm{S - S}}}^{(m,m)}} \right)$ with the $j$th column vector denoted by ${{\bf{\xi }}_{{\rm{S - S}}}^{(m,m,j)}}$. Here, the estimated large-scale fading factors satisfy $\hat \beta _{{\rm{B - M}}}^{(i)} = \frac{{\tau {p_{\tau}}{{\left( {\beta _{{\rm{B - M}}}^{(i)}} \right)}^2}}}{{\tau {p_{\tau}}\beta _{{\rm{B - M}}}^{(i)} + \sigma _0^2}}$, $i\in I$ and $\hat \beta _{{\rm{S - S}}}^{(m,m,j)} = \frac{{\tau {p_{\tau}}{{\left( {\beta _{{\rm{S - S}}}^{(m,m,j)}} \right)}^2}}}{{\tau {p_{\tau}}\sum\limits_{l \in {{\mathcal A}_r}} {\beta _{{\rm{S - S}}}^{(m,l,j)}}  + \sigma _0^2}}$, where $j\in J_m$ and $m\in{\mathcal A}_r$.

\subsection{Data Transmission}
In the downlinks, the received signals at $K$ MUEs and $L$ SUEs in the $m$th small cell are
\begin{equation}\label{sec3:receive_MUE}
{{\bf{y}}_{{\rm{M}}}} = {\bf{G}}_{{\rm{B - M}}}^T{{\bf{W}}_{{\rm{BS}}}}{{\bf{x}}_{{\rm{BS}}}} + \sum\limits_{n = 1}^S {{{\left( {{\bf{G}}_{{\rm{S - M}}}^{(n)}} \right)}^T}{\bf{W}}_{{\rm{SC}}}^{(n)}{\bf{x}}_{{\rm{SC}}}^{(n)}}  + {{\bf{n}}_{{\rm{M}}}}
\end{equation}
\begin{equation}\label{sec3:receive_SUE}
{\bf{y}}_{{\rm{S}}}^{(m)} = {\left( {{\bf{G}}_{{\rm{B - S}}}^{(m)}} \right)^T}{{\bf{W}}_{{\rm{BS}}}}{{\bf{x}}_{{\rm{BS}}}} + \sum\limits_{n = 1}^S {{{\left( {{\bf{G}}_{{\rm{S - S}}}^{(n,m)}} \right)}^T}{\bf{W}}_{{\rm{SC}}}^{(n)}{\bf{x}}_{{\rm{SC}}}^{(n)}}  + {\bf{n}}_{{\rm{S}}}^{(m)}
\end{equation}
respectively, where ${{\bf{W}}_{{\rm{BS}}}}$ and ${{\bf{W}}_{{\rm{SC}}}^{(n)}}$ represent the linear precoding matrices at the BS and $n$th SC, respectively; ${{\bf{x}}_{{\rm{BS}}}} = {\left[ {x_{{\rm{BS}}}^{(1)}, \ldots, x_{{\rm{BS}}}^{(K)}} \right]^T}$ and ${{\bf{x}}_{{\rm{SC}}}^{(n)}} = {\left[ {x_{{\rm{SC}}}^{(n,1)}, \ldots, x_{{\rm{SC}}}^{(n,L)}} \right]^T}$ are the complex-valued data symbols from BS to its MUEs and from $n$th SC to its own SUEs, respectively, satisfying ${\rm {E}}\left[ {{{\bf{x}}_{{\rm{BS}}}}{{\bf{x}}_{{\rm{BS}}}}^H} \right] = {{\bf{I}}_{{N_{{\rm{BS}}}}}}$ and ${\rm {E}}\left[ {{{\bf{x}}_{{\rm{SC}}}^{(n)}}{{\left( {{{\bf{x}}_{{\rm{SC}}}^{(n)}}} \right)}^H}} \right] = {{\bf{I}}_{{N_{{\rm{SC}}}}}}$; and ${{\bf{n}}_{{\rm{M}}}}={\left[ {{{{n}}_{{\rm{M}}}^{(1)}},\ldots, {{{n}}_{{\rm{M}}}^{(K)}}} \right]^T}$ and ${{\bf{n}}_{{\rm{S}}}^{(m)}}={\left[ {{{{n}}_{{\rm{S}}}^{(m,1)}},\ldots, {{{n}}_{{\rm{S}}}^{(m,L)}}} \right]^T}$ involves the AWGN of variance $\sigma_0^2$.

\subsection{MRT Precoding}
Aiming to maximize the received signal-to-noise ratio, the MRT technique is utilized at both the BS and SCs to process the transmit signals towards the corresponding users. Given the estimated channel state information, the MRT precoding is expressed as~\cite{Ngo2014}
\begin{equation}\label{sec3:hat_W1}
{{\bf{W}}_{{\rm{BS}}}} = {{\alpha} _{{\rm{BS}}}}{\hat { \bf G}}_{{\rm{B - M}}}^*,~{\bf{W}}_{{\rm{SC}}}^{(m)} = {\alpha} _{{\rm{SC}}}^{(m)}{\left( {{\hat { \bf G}}_{{\rm{S - S}}}^{(m,m)}} \right)^*}
\end{equation}
where ${{\alpha} _{{\rm{BS}}}}$ and ${\alpha} _{{\rm{SC}}}^{(m)}$ are normalization constants, chosen to satisfy the transmit power constraints at the BS and SCs, respectively. On the basis of (\ref{sec3:hat_W1}) and ${\rm {Tr}}\left\{ {{\bf{AB}}} \right\} = {\rm {Tr}}\left\{ {{\bf{BA}}} \right\}$, we have
\begin{equation}\label{sec3:hat_alpha1}
\begin{split}
{{\alpha} _{{\rm{BS}}}} = \sqrt {\frac{{{p_{{\rm{BS}}}}}}{{{N_{{\rm{BS}}}}{\Phi _{{\rm{B - M}}}}}}},~{\alpha} _{{\rm{SC}}}^{(m)} = \sqrt {\frac{{p_{{\rm{SC}}}^{(m)}}}{{{N_{{\rm{SC}}}}\Phi _{{\rm{S - S}}}^{(m)}}}}
\end{split}
\end{equation}
where ${\Phi _{{\rm{B - M}}}}=\sum\limits_{i = 1}^K {\hat\beta _{{\rm{B - M}}}^{(i)}}$, and $\Phi _{{\rm{S - S}}}^{(m)}{ = }\sum\limits_{l = 1}^L {\hat\beta _{{\rm{S - S}}}^{(m,m,l)}}$ with $m \in\left\{1,\dots,S\right\}$.

\subsection{ZFT Precoding}
Likewise, when ZFT is employed based on imperfect CSI, in which the pseudo-inverse of the estimated channels in (\ref{sec3:estimated_CSI_exp}) are utilized for linear precoding, the precoder is given by~\cite{Ngo2014}
\begin{equation}\label{sec3:hat_W2}
\begin{split}
{{\bf{W}}_{{\rm{BS}}}} &= {{\alpha} _{{\rm{BS}}}}{\hat { \bf G}}_{{\rm{B - M}}}^*{\left( {{{\hat { \bf G}}_{{\rm{B - M}}}^T}{{\hat { \bf G}}_{{\rm{B - M}}}^*}} \right)^{ - 1}} = {{\alpha} _{{\rm{BS}}}}{\hat { \bar {\bf G}}}_{{\rm{B - M}}}^*\\
{\bf{W}}_{{\rm{SC}}}^{(m)} &= {\alpha} _{{\rm{SC}}}^{(m)}{\left( {{\hat { \bf G}}_{{\rm{S - S}}}^{(m,m)}} \right)^*}{\left[ {{\left( {{\hat { \bf G}}_{{\rm{S - S}}}^{(m,m)}} \right)^T}{\left( {{\hat { \bf G}}_{{\rm{S - S}}}^{(m,m)}} \right)^*}} \right]^{ - 1}} \\
&= {\alpha} _{{\rm{SC}}}^{(m)}{\left( {{\hat {\bar{\bf G}}}_{{\rm{S - S}}}^{(m,m)}} \right)^*}
\end{split}
\end{equation}
where ${\hat { \bar {\bf G}}}_{{\rm{B - M}}}={\hat { \bf G}}_{{\rm{B - M}}}\left( {{{\hat { \bf G}}_{{\rm{B - M}}}^H}{{\hat { \bf G}}_{{\rm{B - M}}}}} \right)^{ - 1}$, ${{\hat {\bar{\bf G}}}_{{\rm{S - S}}}^{(m,m)}}= {{\hat { \bf G}}_{{\rm{S - S}}}^{(m,m)}}{\left[ {{\left( {{\hat { \bf G}}_{{\rm{S - S}}}^{(m,m)}} \right)^H} {{\hat { \bf G}}_{{\rm{S - S}}}^{(m,m)}}} \right]^{ - 1}}$, ${{\alpha} _{{\rm{BS}}}}$ and ${\alpha} _{{\rm{SC}}}^{(m)}$ are normalization constants. Similarly, based on (\ref{sec3:hat_W2}) and ${\rm {Tr}}\left\{ {{\bf{AB}}} \right\} = {\rm {Tr}}\left\{ {{\bf{BA}}} \right\}$, we have
\begin{equation}\label{sec3:hat_alpha2}
\begin{split}
{{\alpha} _{{\rm{BS}}}} = \sqrt {\frac{\left( {{N_{{\rm{BS}}}} - K - 1} \right)p_{\rm BS}}{{\Psi }_{\rm{B - M}}}},{\alpha} _{{\rm{SC}}}^{(m)} = \sqrt {\frac{{\left( {{N_{{\rm{SC}}}} - L - 1} \right)p_{{\rm{SC}}}^{(m)}}}{{\Psi _{{\rm{S - S}}}^{(m)}}}}
\end{split}
\end{equation}
where ${{\Psi }_{{\rm{B - M}}}}{\rm{ = }}\sum\limits_{i = 1}^K {\frac{1}{{\hat \beta _{{\rm{B - M}}}^{(i)}}}}$, and $\Psi _{{\rm{S - S}}}^{(m)} = \sum\limits_{l = 1}^L {\frac{1}{{\hat \beta _{{\rm{S - S}}}^{(m,m,l)}}}}$ with $m \in\left\{1,\dots,S\right\}$. The detailed derivation is given in Appendix~\ref{sec3:Appendix_A}.

\section{Achievable Rate Analysis}\label{sec:Achiev_Rate}

The exact rate analysis of the MUE and SUE in the pilot assisted massive MIMO heterogeneous network considered is highly complicated and intractable. In this section, we provide a closed-form capacity lower bound of each user for both MRT and ZFT precoding, respectively. The simple lower bounds can be applied to user scheduling and power allocation optimization as detailed in subsequent sections.

\subsection{MRT Precoding}\label{sec3:MRT}
In practice, only imperfect CSI derived from transmitted pilots is available at each node for linear precoding. Utilizing MRT precoding in (\ref{sec3:receive_MUE}) and (\ref{sec3:receive_SUE}), the received signal of the $i$th~($i\in I$) MUE and $j$th~($j\in J_m$) SUE at the $m$th~($m\in\{1,\dots,S\}$) SC can be rewritten as (\ref{sec3:hat_receive_signal_MUE}) and (\ref{sec3:hat_receive_signal_SUE}).
\begin{table*}
\begin{equation}\label{sec3:hat_receive_signal_MUE}
    \begin{split}
    y_{{\rm{M}}}^{(i)} =& \underbrace{{\alpha _{{\rm{BS}}}}{\left( {\hat {\bf{g}}_{{\rm{B - M}}}^{(i)}} \right)^T}{\left( {\hat {\bf{g}}_{{\rm{B - M}}}^{(i)}} \right)^*}x_{{\rm{BS}}}^{(i)}}_{\text{desired signal}} + \underbrace{{\alpha _{{\rm{BS}}}}{\left( {{{\bf{\xi} }}_{{\rm{B - M}}}^{(i)}} \right)^T}{\left( {\hat {\bf{g}}_{{\rm{B - M}}}^{(i)}} \right)^*}x_{{\rm{BS}}}^{(i)}}_{\text{estimation error induced interference}}+ \underbrace{\sum\limits_{k \ne i}^K {{\alpha _{{\rm{BS}}}}{{\left( {{\bf{g}}_{{\rm{B - M}}}^{(i)}} \right)}^T}{{\left( {\hat {\bf{g}}_{{\rm{B - M}}}^{(k)}} \right)}^*}x_{{\rm{BS}}}^{(k)}}}_{\text{intra-MC interference}}\\
    &~ + \underbrace{\sum\limits_{n = 1}^S {\sum\limits_{l = 1}^L {\alpha _{{\rm{SC}}}^{(n)}{{\left( {{\bf{g}}_{{\rm{S - M}}}^{(n,i)}} \right)}^T}{{\left( {\hat{ \bf{g}}_{{\rm{S - S}}}^{(n,n,l)}} \right)}^*}x_{{\rm{SC}}}^{(n,l)}} }}_{\text{cross-tier interference}}  + \underbrace{n_{{\rm{M}}}^{(i)}}_{\text{noise at MUE}}
    \end{split}
\end{equation}
\begin{equation}\label{sec3:hat_receive_signal_SUE}
    \begin{split}
    y_{{\rm{S}}}^{(m,j)} =& \underbrace{\alpha _{{\rm{SC}}}^{(m)}{\left( {\hat {\bf{g}}_{{\rm{S - S}}}^{(m,m,j)}} \right)^T}{\left( {\hat{\bf{g}}_{{\rm{S - S}}}^{(m,m,j)}} \right)^*}x_{{\rm{SC}}}^{(m,j)}}_{\text{desired signal}} + \underbrace{\alpha _{{\rm{SC}}}^{(m)}{\left( {{\bf{\xi}}_{{\rm{S - S}}}^{(m,m,j)}} \right)^T}{\left( {\hat{\bf{g}}_{{\rm{S - S}}}^{(m,m,j)}} \right)^*}x_{{\rm{SC}}}^{(m,j)}}_{\text{estimation error induced interference}}+ \underbrace{\sum\limits_{i = 1}^K {{\alpha _{{\rm{BS}}}}{{\left( {{\bf{g}}_{{\rm{B - S}}}^{(m,j)}} \right)}^T}{{\left( {\hat{\bf{g}}_{{\rm{B - M}}}^{(i)}} \right)}^*}x_{{\rm{BS}}}^{(i)}}}_{\text{cross-tier interference}} \\
    &+ \underbrace{\sum\limits_{{l_1} \ne j}^L {\alpha _{{\rm{SC}}}^{(m)}{{\left( {{\bf{g}}_{{\rm{S - S}}}^{(m,m,j)}} \right)}^T}{{\left( {\hat{\bf{g}}_{{\rm{S - S}}}^{(m,m,{l_1})}} \right)}^*}x_{{\rm{SC}}}^{(m,{l_1})}}}_{\text{intra-SC interference}} + \underbrace{\sum\limits_{n \ne m}^S {\sum\limits_{{l_2} = 1}^L {\alpha _{{\rm{SC}}}^{(n)}{{\left( {{\bf{g}}_{{\rm{S - S}}}^{(n,m,j)}} \right)}^T}{{\left( {\hat{\bf{g}}_{{\rm{S - S}}}^{(n,n,{l_2})}} \right)}^*}x_{{\rm{SC}}}^{(n,{l_2})}} }}_{\text{inter-SC interference}} + \underbrace{n_{{\rm{S}}}^{(m,j)}}_{\text{noise at SUE}}.
    \end{split}
\end{equation}
\end{table*}
Note that both the BS and small cell nodes treat the estimated channels as the true channels~\cite{Ngo2013}, and the first term is the desired signal. The remaining terms are considered as interferences and noise, including estimation error caused interference term. Accordingly, with imperfect CSI, the ergodic achievable rate of MUE $i$~($i\in I$) and SUE $j$~($j\in J_m$) in the $m$th~($m\in\{1,\dots,S\}$) SC are given by
\begin{equation}\label{sec3:hat_ICSI_MUE_rate}
    \begin{split}
    R_{{\rm{M}}}^{(i)} = {\rm {E}}\left[ {{{\log }_2}\left( {1 + \frac{{\alpha _{{\rm{BS}}}^2{{\left\| {\hat {\bf{g}}_{{\rm{B - M}}}^{(i)}} \right\|}^4}}}{{{\rm{EE}}{{\rm{I}}_i} + {\rm{IM}}{{\rm{I}}_i} + {\rm{CT}}{{\rm{I}}_i} +  \sigma_0^2}}} \right)} \right]
    \end{split}
\end{equation}
\begin{equation}\label{sec3:hat_ICSI_SUE_rate}
    \begin{split}
    &R_{{\rm{S}}}^{(m,j)} = \\
    &{\rm {E}}\left[ {{{\log }_2}\left( {1 + \frac{{{{\left( {\alpha _{{\rm{SC}}}^{(m)}} \right)}^2}{{\left\| {\hat {\bf{g}}_{{\rm{S - S}}}^{(m,m,j)}} \right\|}^4}}}{{{\rm{EE}}{{\rm{I}}_{m,j}} + {\rm{CT}}{{\rm{I}}_{m,j}} + {\rm{IS}}{{\rm{I}}_{m,j}} + {\rm{SS}}{{\rm{I}}_{m,j}} + \sigma_0^2}}} \right)} \right]
    \end{split}
\end{equation}
respectively, where ${\rm{EE}}{{\rm{I}}_i}$, ${\rm{IM}}{{\rm{I}}_i}$ and ${\rm{CT}}{{\rm{I}}_i}$ denote the estimation error induced interference, the intra-MC interference and the cross-tier interference for the $i$th MUE, respectively, given by
\begin{equation}\label{sec3:ICSI_MUE_rate_ext}
\begin{split}
{\rm{EE}}{{\rm{I}}_i} = &\alpha _{{\rm{BS}}}^2{\left| {{{\left( {{\bf{\xi}}_{{\rm{B - M}}}^{(i)}} \right)}^H}{\hat { \bf g}}_{{\rm{B - M}}}^{(i)}} \right|^2},{\rm{IM}}{{\rm{I}}_i} = \sum\limits_{k \ne i}^K {\alpha _{{\rm{BS}}}^2{{\left| {{{\left( {{\bf{g}}_{{\rm{B - M}}}^{(i)}} \right)}^H}{{{\hat{\bf{g}}_{{\rm{B - M}}}^{(k)}} }}} \right|}^2}}\\
&{\rm{CT}}{{\rm{I}}_i} = \sum\limits_{n = 1}^S {\sum\limits_{l = 1}^L {{{\left( {\alpha _{{\rm{SC}}}^{(n)}} \right)}^2}{{\left| {{{\left( {{\bf{g}}_{{\rm{S - M}}}^{(n,i)}} \right)}^H}\hat{\bf{g}}_{{\rm{S - S}}}^{(n,n,l)}} \right|}^2}} },
\end{split}
\end{equation}
and ${\rm{EE}}{{\rm{I}}_{m,j}}$, ${\rm{CT}}{{\rm{I}}_{m,j}}$, ${\rm{IS}}{{\rm{I}}_{m,j}}$ and ${\rm{SS}}{{\rm{I}}_{m,j}}$ denote the estimation error induced interference, the cross-tier interference, the intra-SC interference and the inter-SC interference for the $j$th SUE in the $m$th SC, respectively, given by
\begin{equation}\label{sec3:ICSI_SUE_rate_ext}
\begin{split}
 &{\rm{EE}}{{\rm{I}}_{m,j}}= {\left( {\alpha _{{\rm{SC}}}^{(m)}} \right)^2}{\left| {{{\left( {{\bf{\xi }}_{{\rm{S - S}}}^{(m,m,j)}} \right)}^H}{\hat { \bf g}}_{{\rm{S - S}}}^{(m,m,j)}} \right|^2}\\
 &{\rm{CT}}{{\rm{I}}_{m,j}} = \sum\limits_{i = 1}^K {\alpha _{{\rm{BS}}}^2{{\left| {{{\left( {{\bf{g}}_{{\rm{B - S}}}^{(m,j)}} \right)}^H}{\hat{\bf{g}}_{{\rm{B - M}}}^{(i)}} } \right|}^2}}\\
 &{\rm{IS}}{{\rm{I}}_{m,j}} = \sum\limits_{{l_1} \ne j}^L {{{\left( {\alpha _{{\rm{SC}}}^{(m)}} \right)}^2}{{\left| {{{\left( {{\bf{g}}_{{\rm{S - S}}}^{(m,m,j)}} \right)}^H}{\hat{\bf{g}}_{{\rm{S - S}}}^{(m,m,{l_1})}}} \right|}^2}}\\
 &{\rm{SS}}{{\rm{I}}_{m,j}} = \sum\limits_{n \ne m}^S {\sum\limits_{{l_2} = 1}^L {{{\left( {\alpha _{{\rm{SC}}}^{(n)}} \right)}^2}{{\left| {{{\left( {{\bf{g}}_{{\rm{S - S}}}^{(n,m,j)}} \right)}^H}{\hat{\bf{g}}_{{\rm{S - S}}}^{(n,n,{l_2})}}} \right|}^2}} }.
\end{split}
\end{equation}
Notably, the inter-SC interference ${\rm{SS}}{{\rm{I}}_{m,j}}$ includes the pilot contamination effect caused by pilot reuse. In the above achievable rate expressions, expectations over the estimated instantaneous CSI cannot be further derived into tractable forms. Therefore, we adopt a similar bounding technique of \cite{Ngo2013} to obtain closed form rate expressions, the result of which will provide insights on the impact of different system parameters and facilitate further optimizations.

By the convexity of ${\log _2}\left( {1 + \frac{1}{x}} \right)$ and Jensen's inequality, from (\ref{sec3:hat_ICSI_MUE_rate}) and (\ref{sec3:hat_ICSI_SUE_rate}), a lower bound on the achievable rate is obtained as
\begin{equation}\label{sec3:hat_SCSI_MUE_rate2}
    \begin{split}
    R_{{\rm{0,M,M}}}^{(i)} = {\log _2}\left( {1 + {{\left( {{\rm {E}}\left[ {\frac{{{\rm{EE}}{{\rm{I}}_i} + {\rm{IM}}{{\rm{I}}_i} + {\rm{CT}}{{\rm{I}}_i} + \sigma _0^2}}{{\alpha _{{\rm{BS}}}^2{{\left\| {\hat {\bf{g}}_{{\rm{B - M}}}^{(i)}} \right\|}^4}}}} \right]} \right)}^{ - 1}}} \right)
    \end{split}
\end{equation}
\begin{equation}\label{sec3:hat_SCSI_SUE_rate2}
    \begin{split}
    &R_{{\rm{0,S,M}}}^{(m,j)} = {\log _2}\left( {1 + }\right.\\
    &\left.{{{\left( {{\rm {E}}\left[ {\frac{{{\rm{EE}}{{\rm{I}}_{m,j}} + {\rm{CT}}{{\rm{I}}_{m,j}} + {\rm{IS}}{{\rm{I}}_{m,j}} + {\rm{SS}}{{\rm{I}}_{m,j}} + \sigma _0^2}}{{{{\left( {\alpha _{{\rm{SC}}}^{(m)}} \right)}^2}{{\left\| {\hat {\bf{g}}_{{\rm{S - S}}}^{(m,m,j)}} \right\|}^4}}}} \right]} \right)}^{ - 1}}} \right).
    \end{split}
\end{equation}

\emph{Theorem 1:} With imperfect CSI based MRT, $N_{\rm BS}\ge 2$ and $N_{\rm SC}\ge 2$, the downlink achievable rate of the $i$th~($i\in I$) MUE and $j$th~($j\in J_m$) SUE in the $m$th~($m\in\{1,\dots,S\}$) SC, for finite $N_{\rm BS}$ and $N_{\rm SC}$, are lower bounded by
\begin{equation}\label{sec3:hat_SCSI_MUE_rate2_CF}
    \begin{split}
    R_{{\rm{0,M,M}}}^{(i)} = {\log _2}\left( {1 + \frac{{a_{\rm MR}^{(i)}p_{{\rm{BS}}}}}{{b_{\rm MR}^{(i)}p_{{\rm{BS}}} + \sum\limits_{n = 1}^S {{c_{\rm MR}^{(n,i)}}p_{\rm SC}^{(n)}} + \sigma _0^2}}} \right)
    \end{split}
\end{equation}
\begin{equation}\label{sec3:hat_SCSI_SUE_rate2_CF}
    \begin{split}
    R_{{\rm{0,S,M}}}^{(m,j)} = {\log _2}\left( {1 + \frac{{{d_{\rm MR}^{(m,j)}}}p_{\rm SC}^{(m)}}{{\sum\limits_{n = 1}^S {{e_{\rm MR}^{(n,m,j)}}p_{\rm SC}^{(n)}}+ f_{\rm MR}^{(m,j)}p_{{\rm{BS}}} + \sigma _0^2}}} \right)
    \end{split}
\end{equation}
where $a_{\rm M}^{(i)} = \frac{({N_{{\rm{BS}}}} - 1)({N_{{\rm{BS}}}} - 2){{\left( {\hat \beta _{{\rm{B - M}}}^{(i)}} \right)}^2}}{{{N_{{\rm{BS}}}}{\Phi _{{\rm{B - M}}}}}}$, $b_{\rm M}^{(i)} = \beta _{{\rm{B - M}}}^{(i)} - \frac{2}{{{N_{{\rm{BS}}}}}}\hat \beta _{{\rm{B - M}}}^{(i)} - \frac{{\left( {{N_{{\rm{BS}}}} - 4} \right){{\left( {\hat \beta _{{\rm{B - M}}}^{(i)}} \right)}^2}{\rm{ + }}2\beta _{{\rm{B - M}}}^{(i)}\hat \beta _{{\rm{B - M}}}^{(i)}}}{{{N_{{\rm{BS}}}}{{\Phi }_{{\rm{B - M}}}}}}$, $c_{\rm M}^{(n,i)} = \beta _{{\rm{S - M}}}^{(n,i)}$, $d_{\rm M}^{(m,j)} = \frac{\left( {{N_{{\rm{SC}}}} - 1} \right)\left( {{N_{{\rm{SC}}}} - 2} \right){{\left( {\hat \beta _{{\rm{S - S}}}^{(m,m,j)}} \right)}^2}}{{{N_{{\rm{SC}}}}\Phi _{{\rm{S - S}}}^{(m)}}}$, $e_{\rm M}^{(n,m,j)} = \left\{\begin{array}{l}
    \beta _{{\rm{S - S}}}^{(n,m,j)},~n \ne m,~n \notin {\mathcal A}_r\\
    \frac{{N_{\rm{SC}}}\left({\beta _{{\rm{S - S}}}^{(n,m,j)}}\right)^2\left({\hat \beta _{{\rm{S - S}}}^{(n,n,j)}}\right)^2}{\left({\beta _{{\rm{S - S}}}^{(n,n,j)}}\right)^2\Phi _{\rm{S - S}}^{(n)}}+\beta _{{\rm{S - S}}}^{(n,m,j)},~n \ne m,~n \in {\mathcal A}_r\\
    \beta _{{\rm{S - S}}}^{(m,m,j)} - \frac{2}{{{N_{{\rm{SC}}}}}}\hat \beta _{{\rm{S - S}}}^{(m,m,j)} - \\
    \frac{{\left( {{N_{{\rm{SC}}}} - 4} \right){{\left( {\hat \beta _{{\rm{S - S}}}^{(m,m,j)}} \right)}^2} + 2\beta _{{\rm{S - S}}}^{(m,m,j)}\hat \beta _{{\rm{S - S}}}^{(m,m,j)}}}{{{N_{{\rm{SC}}}}\Phi _{{\rm{S - S}}}^{(m)}}},~n = m
\end{array}\right.$ and $f_{\rm M}^{(m,j)} = \beta _{{\rm{B - S}}}^{(m,j)}$.

\emph{Proof:} See Appendix~\ref{sec3:Appendix_B}.

\emph{Remark 1:} Since the proof in Appendix~\ref{sec3:Appendix_B} does not use any asymptotic assumptions on the antenna size, Theorem 1 is also valid for conventional scale MIMO systems. The capacity lower bounds for perfect CSI can be obtained by setting $\hat\beta_{\cdot}^{(\cdot)} = \beta_{\cdot}^{(\cdot)}$ in (\ref{sec3:hat_SCSI_MUE_rate2_CF}) and (\ref{sec3:hat_SCSI_SUE_rate2_CF}). Moreover, it can be observed from Appendix~\ref{sec3:Appendix_B} that all the interferences contained in the received signals of the MUEs (i.e., EEI, IMI and CTI) can be significantly mitigated relative to the desired signals by increasing $N_{\rm BS}$. Similarly, all the interference effect at the SUEs (i.e., EEI, CTI, ISI and SSI) is able to be reduced by increasing $N_{\rm SC}$. These observations support the use of large scale antenna arrays at both BS and SC. In addition, the expression of $R_{{\rm{0,M,M}}}^{(i)}$ indicates that the MUE rate increases monotonically with $N_{\rm BS}$ but has no relationship with $N_{\rm SC}$. Similarly, the SUE rate increases monotonically with $N_{\rm SC}$ and is independent of $N_{\rm BS}$.

\subsection{ZFT Precoding}\label{sec3:ZFT}
For imperfect CSI based ZFT precoding, the received signal can be rewritten as (\ref{sec3:hat_receive_signal_MUE2}) and (\ref{sec3:hat_receive_signal_SUE2}),
\begin{table*}
\begin{equation}\label{sec3:hat_receive_signal_MUE2}
    \begin{split}
    y_{{\rm{M}}}^{(i)} = &\underbrace{{\alpha _{{\rm{BS}}}}x_{{\rm{BS}}}^{(i)}}_{\text{desired signal}} + \underbrace{{\alpha _{{\rm{BS}}}}{\left( {{\bf {\xi }}_{{\rm{B - M}}}^{(i)}} \right)^T}{\left( {\hat {\bar {\bf g}}}_{{\rm{B - M}}}^{(i)} \right)^*}x_{{\rm{BS}}}^{(i)}}_{\text{estimation error induced interference}} + \underbrace{\sum\limits_{k \ne i}^K {{\alpha _{{\rm{BS}}}}{{\left( {{\bf{\xi}}_{{\rm{B - M}}}^{(i)}} \right)}^T}{{\left( {{\hat {\bar {\bf g}}}_{{\rm{B - M}}}^{(k)}} \right)}^*}x_{{\rm{BS}}}^{(k)}}}_{\text{intra-MC interference}}\\
     &~+\underbrace{\sum\limits_{n = 1}^S {\sum\limits_{l = 1}^L {\alpha _{{\rm{SC}}}^{(n)}{{\left( {{\bf{g}}_{{\rm{S - M}}}^{(n,i)}} \right)}^T}{{\left( {\hat{\bar {\bf g}}}_{{\rm{S - S}}}^{(n,n,l)} \right)}^*}x_{{\rm{SC}}}^{(n,l)}} }}_{\text{cross-tier interference}}  + \underbrace{n_{{\rm{M}}}^{(i)}}_{\text{noise at MUE}}
    \end{split}
\end{equation}
\begin{equation}\label{sec3:hat_receive_signal_SUE2}
    \begin{split}
    y_{{\rm{S}}}^{(m,j)} &= \underbrace{\alpha _{{\rm{SC}}}^{(m)}x_{{\rm{SC}}}^{(m,j)}}_{\text{desired signal}} + \underbrace{\alpha _{{\rm{SC}}}^{(m)}{\left( {{\bf{\xi}}_{{\rm{S - S}}}^{(m,m,j)}} \right)^T}{\left( {\hat {\bar{\bf g}}}_{{\rm{S - S}}}^{(m,m,j)} \right)^*}x_{{\rm{SC}}}^{(m,j)}}_{\text{estimation error induced interference}} + \underbrace{\sum\limits_{i = 1}^K {\alpha _{{\rm{BS}}}}{{\left( {{\bf{g}}_{{\rm{B - S}}}^{(m,j)}} \right)}^T}{{\left( {\hat {\bar{\bf g}}}_{{\rm{B - M}}}^{(i)} \right)^*}x_{{\rm{BS}}}^{(i)}}}_{\text{cross-tier interference}} \\
    &~+ \underbrace{\sum\limits_{{l_1} \ne j}^L {\alpha _{{\rm{SC}}}^{(m)}{{\left( {{\bf{\xi}}_{{\rm{S - S}}}^{(m,m,j)}} \right)}^T}{{\left( {{\hat {\bar{\bf g}}}_{{\rm{S - S}}}^{(m,m,{l_1})}} \right)}^*}x_{{\rm{SC}}}^{(m,{l_1})}}}_{\text{intra-SC interference}}+ \underbrace{\sum\limits_{n \ne m}^S {\sum\limits_{{l_2} = 1}^L \alpha _{{\rm{SC}}}^{(n)}{{\left( {{\bf{g}}_{{\rm{S - S}}}^{(n,m,j)}} \right)}^T}{{\left( {\hat {\bar{\bf g}}}_{{\rm{S - S}}}^{(n,n,{l_2})} \right)}^*}x_{{\rm{SC}}}^{(n,{l_2})} }}_{\text{inter-SC interference}} + \underbrace{n_{{\rm{S}}}^{(m,j)}}_{\text{noise at SUE}}
    \end{split}
\end{equation}
\end{table*}
respectively, where ${\left( {\hat {\bf{g}}_{{\rm{B - M}}}^{(i)}} \right)^T}{\left( {\hat {\bar{\bf g}}}_{{\rm{B - M}}}^{(i)} \right)^*}=1$, ${\left( {\hat {\bf{g}}_{{\rm{S - S}}}^{(m,m,j)}} \right)^T}{\left( {\hat {\bar{\bf g}}}_{{\rm{S - S}}}^{(m,m,j)} \right)^*}=1$, and the intra-MC interference ${\rm{IM}}{{\rm{I}}_i}$ and intra-SC interference ${\rm{IS}}{{\rm{I}}_{m,j}}$ are reduced because ZFT precoding is able to null multi-user interference signals, i.e., ${{{\left( {\hat{\bf{g}}_{{\rm{B - M}}}^{(i)}} \right)}^T}{{\left( {{\hat {\bar{\bf g}}}_{{\rm{B - M}}}^{(k)}} \right)}^*}}=0$ and ${{{\left( {\hat{\bf{g}}_{{\rm{S - S}}}^{(m,m,j)}} \right)}^T}{{\left( {{\hat {\bar{\bf g}}}_{{\rm{S - S}}}^{(m,m,{l})}} \right)}^*}}=0$ for $\forall k\ne i$ and $\forall l \ne j$. Similarly, by the convexity of ${\log _2}\left( {1 + \frac{1}{x}} \right)$ and Jensen's inequality, capacity lower bounds of MUE $i$~($i\in I$) and SUE $j$~($j\in J_m$) in the $m$th~($m\in\{1,\dots,S\}$) SC in (\ref{sec3:hat_SCSI_MUE_rate2}) and (\ref{sec3:hat_SCSI_SUE_rate2}) become
\begin{equation}\label{sec3:hat_SCSI_MUE_rate22}
    \begin{split}
    R_{{\rm{0,M,Z}}}^{(i)} = {\log _2}\left( {1 + {{\left( {{\rm {E}}\left[ {\frac{{{\rm{EE}}{{\rm{I}}_i} + {\rm{IM}}{{\rm{I}}_i} + {\rm{CT}}{{\rm{I}}_i} + \sigma _0^2}}{{\alpha _{{\rm{BS}}}^2}}} \right]} \right)}^{ - 1}}} \right)
    \end{split}
\end{equation}
\begin{equation}\label{sec3:hat_SCSI_SUE_rate22}
    \begin{split}
    &R_{{\rm{0,S,Z}}}^{(m,j)} = {\log _2}\left( {1 + }\right.\\
    &\left.{{{\left( {{\rm {E}}\left[ {\frac{{{\rm{EE}}{{\rm{I}}_{m,j}} + {\rm{CT}}{{\rm{I}}_{m,j}} + {\rm{IS}}{{\rm{I}}_{m,j}} + {\rm{SS}}{{\rm{I}}_{m,j}} + \sigma _0^2}}{{{{\left( {\alpha _{{\rm{SC}}}^{(m)}} \right)}^2}}}} \right]} \right)}^{ - 1}}} \right).
    \end{split}
\end{equation}
where (\ref{sec3:ICSI_MUE_rate_ext2}) and (\ref{sec3:ICSI_SUE_rate_ext2}).
\begin{table*}
\begin{equation}\label{sec3:ICSI_MUE_rate_ext2}
\begin{split}
&{\rm{EE}}{{\rm{I}}_i} = \alpha _{{\rm{BS}}}^2{\left| {{{\left( {{\bf{\xi}}_{{\rm{B - M}}}^{(i)}} \right)}^H}{\hat {\bar{\bf g}}}_{{\rm{B - M}}}^{(i)}} \right|^2},~{\rm{IM}}{{\rm{I}}_i} = \sum\limits_{k \ne i}^K {\alpha _{{\rm{BS}}}^2{{\left| {{{\left( {{\bf{\xi}}_{{\rm{B - M}}}^{(i)}} \right)}^H}{{{\hat {\bar{\bf g}}}_{{\rm{B - M}}}^{(k)}}}} \right|}^2}}\\
&{\rm{CT}}{{\rm{I}}_i} = \sum\limits_{n = 1}^S {\sum\limits_{l = 1}^L {{{\left( {\alpha _{{\rm{SC}}}^{(n)}} \right)}^2}{{\left| {{{\left( {{\bf{g}}_{{\rm{S - M}}}^{(n,i)}} \right)}^H}{{{\hat {\bar{\bf g}}}_{{\rm{S - S}}}^{(n,n,l)}}}} \right|}^2}} }
\end{split}
\end{equation}
\begin{equation}\label{sec3:ICSI_SUE_rate_ext2}
\begin{split}
&{\rm{EE}}{{\rm{I}}_{m,j}} = {\left( {\alpha _{{\rm{SC}}}^{(m)}} \right)^2}{\left| {{{\left( {{\bf{\xi }}_{{\rm{S - S}}}^{(m,m,j)}} \right)}^H}{\hat {\bar{\bf g}}}_{{\rm{S - S}}}^{(m,m,j)}} \right|^2},~{\rm{CT}}{{\rm{I}}_{m,j}} = \sum\limits_{i = 1}^K {\alpha _{{\rm{BS}}}^2{{\left| {{{\left( {{\bf{g}}_{{\rm{B - S}}}^{(m,j)}} \right)}^H}{{\hat {\bar{\bf g}}}_{{\rm{B - M}}}^{(i)}}} \right|}^2}}\\
&{\rm{IS}}{{\rm{I}}_{m,j}} = \sum\limits_{{l_1} \ne j}^L {{{\left( {\alpha _{{\rm{SC}}}^{(m)}} \right)}^2}{{\left| {{{\left( {{\bf{\xi}}_{{\rm{S - S}}}^{(m,m,j)}} \right)}^H}{{\hat {\bar{\bf g}}}_{{\rm{S - S}}}^{(m,m,{l_1})}}} \right|}^2}},~{\rm{SS}}{{\rm{I}}_{m,j}} = \sum\limits_{n \ne m}^S {\sum\limits_{{l_2} = 1}^L {{{\left( {\alpha _{{\rm{SC}}}^{(n)}} \right)}^2}{{\left| {{{\left( {{\bf{g}}_{{\rm{S - S}}}^{(n,m,j)}} \right)}^H}{{\hat {\bar{\bf g}}}_{{\rm{S - S}}}^{(n,n,{l_2})}}} \right|}^2}} }.
\end{split}
\end{equation}
\end{table*}

\emph{Theorem 2:} With imperfect CSI based ZFT, $N_{\rm BS}\ge 2$ and $N_{\rm SC}\ge 2$, the downlink achievable rate of the $i$th~($i\in I$) MUE and $j$th~($j\in J_m$) SUE in the $m$th~($m\in\{1,\dots,S\}$) SC, for finite $N_{\rm BS}$ and $N_{\rm SC}$, are lower bounded by
\begin{equation}\label{sec3:hat_SCSI_MUE_rate2_CF2}
    \begin{split}
    R_{{\rm{0,M,Z}}}^{(i)} = {\log _2}\left( {1 + \frac{{a_{\rm ZF}^{(i)}p_{{\rm{BS}}}}}{{b_{\rm ZF}^{(i)}p_{{\rm{BS}}} + \sum\limits_{n = 1}^S {{c_{\rm ZF}^{(n,i)}}p_{\rm SC}^{(n)}} + \sigma _0^2}}} \right)
    \end{split}
\end{equation}
\begin{equation}\label{sec3:hat_SCSI_SUE_rate2_CF2}
    \begin{split}
    R_{{\rm{0,S,Z}}}^{(m,j)} = {\log _2}\left( {1 + \frac{{{d_{\rm ZF}^{(m,j)}}}p_{\rm SC}^{(m)}}{{\sum\limits_{n = 1}^S {{e_{\rm ZF}^{(n,m,j)}}p_{\rm SC}^{(n)}}+ f_{\rm ZF}^{(m,j)}p_{{\rm{BS}}} + \sigma _0^2}}} \right)
    \end{split}
\end{equation}
where $a_{\rm Z}^{(i)} = \frac{{{N_{{\rm{BS}}}} - K - 1}}{{{\Psi _{{\rm{B - M}}}}}}$, $b_{\rm Z}^{(i)} = \xi _{{\rm{B - M}}}^{(i)}$, $c_{\rm Z}^{(n,i)} = \beta _{{\rm{S - M}}}^{(n,i)}$, $d_{\rm Z}^{(m,j)} = \frac{{{N_{{\rm{SC}}}} - L - 1}}{{\Psi _{{\rm{S - S}}}^{(m)}}}$, $f_{\rm Z}^{(m,j)} = \beta _{{\rm{B - S}}}^{(m,j)}$, and $e_{\rm Z}^{(n,m,j)} = \left\{\begin{array}{l}
    \beta _{{\rm{S - S}}}^{(n,m,j)},~n \ne m,~n \notin {\mathcal A}_r\\
    {\beta _{{\rm{S - S}}}^{(n,m,j)}}-\frac{{\beta _{{\rm{S - S}}}^{(n,m,j)}}}{{\Psi _{{\rm{S - S}}}^{(n)}}{\hat\beta _{{\rm{S - S}}}^{(n,n,j)}}} + \frac{\left(N_{\rm{SC}} - L - 1\right)\left({\beta _{{\rm{S - S}}}^{(n,m,j)}}\right)^2}{{\Psi _{{\rm{S - S}}}^{(n)}}\left({\beta _{{\rm{S - S}}}^{(n,n,j)}}\right)^2},n \ne m,n \in {\mathcal A}_r\\
    \xi _{{\rm{S - S}}}^{(m,m,j)},~n = m.
    \end{array}\right.$

\emph{Proof:} See Appendix~\ref{sec3:Appendix_D}.

\emph{Remark 2:} Similar to MRT, conclusions in Remark 1 are also valid for ZFT based Theorem 2. Moreover, the capacity lower bounds in (\ref{sec3:hat_SCSI_MUE_rate2_CF2}) and (\ref{sec3:hat_SCSI_SUE_rate2_CF2}) indicate that $R_{{\rm{0,M,ZF}}}^{(i)}$ decreases monotonically as the estimation error $\xi _{{\rm{B - M}}}^{(i)}$ increases for fixed ${\Psi _{{\rm{B - M}}}}$. Similar conclusions are drawn for the capacity lower bounds of SUEs.

\emph{Remark 3:} From (\ref{sec3:hat_SCSI_MUE_rate2_CF2}), we can conclude that the expression of $R_{{\rm{0,M,Z}}}^{(i)}$ for the $i$th MUE involves only $N_{\rm BS}$, but no $N_{\rm SC}$, which indicates that the capacity lower bound of MUE depends on $N_{\rm BS}$ but has no relationship with $N_{\rm SC}$. Furthermore, the signal-to-interference-plus-noise ratio (SINR) involved in $R_{{\rm{0,M,Z}}}^{(i)}$ is approximately a linearly increasing function of $N_{\rm BS}$, when $N_{\rm BS} \gg K$. Similarly, from (\ref{sec3:hat_SCSI_SUE_rate2_CF2}), it can be concluded that the capacity lower bound of SUE increases monotonically with $N_{\rm SC}$ but is independent of $N_{\rm BS}$. However, due to the pilot contamination effect caused by pilot reuse, the increase of the SINR in $R_{{\rm{0,S,Z}}}^{(m,j)}$ is not linear even when $N_{\rm{SC}} \gg L$.

\subsection{Asymptotic Analysis with Massive Arrays}\label{asymptotic}
Having obtained the closed-form expressions for the achievable rate in (\ref{sec3:hat_SCSI_MUE_rate2_CF}) and (\ref{sec3:hat_SCSI_SUE_rate2_CF}), this subsection provides the asymptotic analysis under two different cases when the number of antennas approaches infinity. Suppose that all SCs have the same transmit power, i.e., $p_{\text SC}^{(1)} = \cdots  = p_{\text SC}^{(S)} = p_{\text SC}$, and $N_{\rm BS} = \lambda N_{\rm SC}$ with $\lambda \ge 10$.

\emph{Proposition 1:} In case I where $p_{\tau}$ is fixed, $p_{\rm SC}^{(s)} = p_{\rm SC} = \frac{{{E_{\rm{SC}}}}}{N_{\rm SC}^{\chi_1}}$~($s = 1,\cdots,S$), ${p_{\rm{BS}}} = \frac{{{E_{\rm{BS}}}}}{N_{\rm BS}^{\eta_1}}$, and $E_{\rm{SC}}$ and $E_{\rm{BS}}$ are fixed, to achieve non-vanishing user rate as $N_{\rm SC} \to \infty $ with $N_{\rm BS} = \lambda N_{\rm SC}$, the SC and BS transmit power scaling factors $\chi_1$ and $\eta_1$ must satisfy $0\le \chi_1 \le 1$ and $0 \le \eta_1 \le 1$. When $\chi_1 = \eta_1 = 1$, the asymptotic achievable rate expressions of the $i$th~($i\in I$) MUE and $j$th~($j\in J_m$) SUE in the $m$th~($m\in\{1,\dots,S\}$) SC for imperfect CSI based MRT and ZFT are (\ref{sec3:hat_SCSI_MUE_rate_P1}) and (\ref{sec3:hat_SCSI_SUE_rate_P1}),
\begin{table*}
\begin{equation}\label{sec3:hat_SCSI_MUE_rate_P1}
R_{{\rm{0,M,M}}}^{(i)}\mathop  {\longrightarrow} \limits_{{N_{\rm SC}} \to \infty }^{a.s.} {\log _2}\left( {1 + \frac{{{E_{{\rm{BS}}}}{{\left( {\hat \beta _{{\rm{B - M}}}^{(i)}} \right)}^2}}}{{{{\Phi }_{{\rm{B - M}}}}\sigma _0^2}}} \right),R_{{\rm{0,S,M}}}^{(m,j)}\mathop  {\longrightarrow} \limits_{{N_{\rm SC}} \to \infty }^{a.s.} {\log _2}\left( {1 + \frac{{E_{{\rm{SC}}}}{{\left( {\hat \beta _{{\rm{S - S}}}^{(m,m,j)}} \right)}^2}/{\Phi _{{\rm{S - S}}}^{(m)}}}{\sigma _0^2 + \sum\limits_{\scriptstyle n \ne m\hfill\atop
\scriptstyle n \in {{\mathcal A}_r}\hfill} \frac{{E_{{\rm{SC}}}}\left({\hat \beta _{{\rm{S - S}}}^{(n,n,j)}}{\beta _{{\rm{S - S}}}^{(n,m,j)}}\right)^2}{\Phi _{\rm{S - S}}^{(n)}\left({\beta _{{\rm{S - S}}}^{(n,n,j)}}\right)^2}}} \right)
\end{equation}
\begin{equation}\label{sec3:hat_SCSI_SUE_rate_P1}
R_{{\rm{0,M,Z}}}^{(i)}\mathop  {\longrightarrow} \limits_{{N_{\rm SC}} \to \infty }^{a.s.} {\log _2}\left( {1 + \frac{{{E_{{\rm{BS}}}}}}{{{{\Psi }_{{\rm{B - M}}}}\sigma _0^2}}} \right),R_{{\rm{0,S,Z}}}^{(m,j)}\mathop  {\longrightarrow} \limits_{{N_{\rm SC}} \to \infty }^{a.s.} {\log _2}\left( {1 + \frac{E_{\rm{SC}}/\Psi _{{\rm{S - S}}}^{(m)}}{{\sigma _0^2+\sum\limits_{\scriptstyle n \ne m\hfill\atop
\scriptstyle n \in {{\mathcal A}_r}\hfill} \frac{{E_{{\rm{SC}}}}\left(\beta_{{\rm{S - S}}}^{(n,m,j)}\right)^2}{\Psi _{\rm{S - S}}^{(n)}\left(\beta_{{\rm{S - S}}}^{(n,n,j)}\right)^2}}}} \right)
\end{equation}
\begin{equation}\label{sec3:hat_SCSI_MUE_rate_P2}
\begin{split}
&R_{{\rm{0,M,M}}}^{(i)}\mathop  {\longrightarrow} \limits_{{N_{\rm SC}} \to \infty }^{a.s.} {\log _2}\left( {1 + \frac{{\lambda^{\theta}  \tau {E_\tau }{E_{\rm BS }}{{\left( {\beta _{{\rm{B - M}}}^{(i)}} \right)}^4}}}{{{{\sum\limits_{k = 1}^K {\left( {\beta _{{\rm{B - M}}}^{(k)}} \right)}^2}}\sigma _0^4}}} \right),R_{{\rm{0,S,M}}}^{(m,j)}\mathop  {\longrightarrow} \limits_{{N_{\rm SC}} \to \infty }^{a.s.} {\log_2}\left( {1 + \frac{{\tau {E_\tau }{E_{\rm SC}}{{\left( {\beta _{{\rm{S - S}}}^{(m,m,j)}} \right)}^4}}}{{{{\sum\limits_{l = 1}^L {\left( {\beta _{{\rm{S - S}}}^{(m,m,l)}} \right)}^2 }}\sigma _0^4}}} \right)
\end{split}
\end{equation}
\begin{equation}\label{sec3:hat_SCSI_SUE_rate_P2}
\begin{split}
&R_{{\rm{0,M,Z}}}^{(i)}\mathop  {\longrightarrow} \limits_{{N_{\rm SC}} \to \infty }^{a.s.} {\log _2}\left( {1 + \frac{\lambda^{\theta}  \tau {E_\tau }{E_{\rm BS}}}{{\sum\limits_{k = 1}^K {\left( {\beta _{{\rm{B - M}}}^{(k)}} \right)}^{-2}}\sigma _0^4}} \right),R_{{\rm{0,S,Z}}}^{(m,j)}\mathop  {\longrightarrow} \limits_{{N_{\rm SC}} \to \infty }^{a.s.} {\log_2}\left( {1 + \frac{{\tau {E_\tau }}{E_{\rm SC}}}{{{{\sum\limits_{l = 1}^L {\left( {\beta _{{\rm{S - S}}}^{(m,m,l)}} \right)}^{-2} }}\sigma _0^4}}} \right)
\end{split}
\end{equation}
\end{table*}
respectively, which show that the transmit powers at both BS and SCs can be scaled down by up to $\frac{1}{{N_{\rm SC}}}$ to maintain a given rate in case I. When $0\le \chi_1 < 1$ and $0 \le \eta_1 < 1$, the asymptotic achievable rate of each user approaches to infinity as ${N_{\rm SC}} \to \infty$.

\emph{Remark 4:} Obviously, when the pilot reuse factor $\gamma = S$, i.e., no pilot reuse, we have $R_{{\rm{0,S,M}}}^{(m,j)}\mathop  {\longrightarrow} \limits^{a.s.} {\log _2}\left( {1 + \frac{{E_{{\rm{SC}}}}{{\left( {\hat \beta _{{\rm{S - S}}}^{(m,m,j)}} \right)}^2}}{{\Phi _{{\rm{S - S}}}^{(m)}}\sigma _0^2 }} \right)$ and $R_{{\rm{0,S,Z}}}^{(m,j)}\mathop  {\longrightarrow} \limits^{a.s.} {\log _2}\left( {1 + \frac{E_{\rm{SC}}}{{\Psi _{{\rm{S - S}}}^{(m)}\sigma _0^2}}} \right)$. For practical system configurations, we suppose that $\chi_1\ge \eta_1$ in case I to guarantee $p_{\rm SC}<p_{\rm BS}$. When $0< \chi_1 < 1$ and $0 < \eta_1 < 1$, the asymptotic achievable rate of both MUEs and SUEs approaches to infinity as ${N_{\rm SC}} \to \infty$ for both MRT and ZFT. When $\chi_1=1$ and $\eta_1=0$, we have $R_{{\rm{0,M,M}}}^{(i)}\mathop  {\longrightarrow} \limits^{a.s.} \infty$, $R_{{\rm{0,M,Z}}}^{(i)}\mathop  {\longrightarrow} \limits^{a.s.} \infty$, $R_{{\rm{0,S,M}}}^{(m,j)}\mathop  {\longrightarrow} \limits^{a.s.} {\log _2}\left( {1 + \frac{{\frac{{E_{{\rm{SC}}}}{{\left( {\hat \beta _{{\rm{S - S}}}^{(m,m,j)}} \right)}^2}}{\Phi _{{\rm{S - S}}}^{(m)}E_{\rm BS}}}}{{\beta_{\rm B-S}^{(m,j)}+\sum\limits_{\scriptstyle n \ne m\hfill\atop
\scriptstyle n \in {{\mathcal A}_r}\hfill} \frac{{E_{{\rm{SC}}}}\left({\hat \beta _{{\rm{S - S}}}^{(n,n,j)}}{\beta _{{\rm{S - S}}}^{(n,m,j)}}\right)^2}{E_{\rm BS}\Phi _{\rm{S - S}}^{(n)}\left({\beta _{{\rm{S - S}}}^{(n,n,j)}}\right)^2}+\frac{\sigma _0^2}{E_{\rm BS}}}}} \right)$, and $R_{{\rm{0,S,Z}}}^{(m,j)}\mathop  {\longrightarrow} \limits^{a.s.} {\log _2}\left( {1 + \frac{E_{\rm{SC}}/\Psi _{{\rm{S - S}}}^{(m)}}{{\beta_{\rm B-S}^{(m,j)}E_{\rm BS}+\sum\limits_{\scriptstyle n \ne m\hfill\atop
\scriptstyle n \in {{\mathcal A}_r}\hfill} \frac{{E_{{\rm{SC}}}}\left(\beta_{{\rm{S - S}}}^{(n,m,j)}\right)^2}{\Psi _{\rm{S - S}}^{(n)}\left(\beta_{{\rm{S - S}}}^{(n,n,j)}\right)^2}+\sigma _0^2}}} \right)$, respectively, indicating that the cross-tier interferences at SUEs can not be eliminated when $p_{\rm SC}$ is scaled down proportionally to $\frac{1}{{N_{\rm SC}}}$ with fixed ${p_{\rm{BS}}}$ in case I.

\emph{Proposition 2:} In case II where $p_{\tau}=\frac{{{E_{\tau}}}}{N_{\rm SC}^{\theta}}$, $p_{\rm SC}^{(s)} = p_{\rm SC} = \frac{{{E_{\rm{SC}}}}}{N_{\rm SC}^{\chi_2}}$~($s = 1,\cdots,S$), ${p_{\rm{BS}}} = \frac{{{E_{\rm{BS}}}}}{N_{\rm BS}^{\eta_2}}$, and ${E_{\tau}}$, $E_{\rm{BS}}$ and $E_{\rm{SC}}$ are fixed, to achieve non-vanishing user rate as $N_{\rm SC} \to \infty$ with $N_{\rm BS} = \lambda N_{\rm SC}$ and the pilot reuse factor $\gamma = S$, the pilot, SC and BS transmit power scaling factors $\theta$, $\chi_2$ and $\eta_2$ must satisfy $0<\theta\le 1$, $0\le \chi_2 \le 1-\theta$ and $0 \le \eta_2 \le 1-\theta$. When $0<\theta<1$ and $\chi_2 = \eta_2 = 1 - \theta$, the asymptotic achievable rate expressions of the $i$th~($i\in I$) MUE and $j$th~($j\in J_m$) SUE in the $m$th~($m\in\{1,\dots,S\}$) SC for imperfect CSI based MRT and ZFT are (\ref{sec3:hat_SCSI_MUE_rate_P2}) and (\ref{sec3:hat_SCSI_SUE_rate_P2}), respectively, from which we conclude that the transmit powers of BS and SCs can only be reduced by up to $\frac{1}{N_{\rm SC}^{1-\theta}}$ with the pilot transmit power set as $p_{\tau}=\frac{{{E_{\tau}}}}{N_{\rm SC}^{\theta}}$ and pilot reuse factor $\gamma = S$~(no pilot reuse) in case II. When $0<\theta< 1$, $0\le \chi_2 < 1-\theta$ and $0 \le \eta_2 < 1-\theta$, the asymptotic achievable rate of each user approaches to infinity as ${N_{\rm SC}} \to \infty$.

\emph{Remark 5:} To guarantee MUE and SUE achievable rate, $\chi_2=\eta_2=0$ should be satisfied in case II when $\theta=1$, which means that the pilot power can be scaled down by up to $\frac{1}{N_{\rm SC}}$ with fixed transmit power at both BS and SC nodes. Then, the asymptotic achievable rate can be expressed as $R_{{\rm{0,M,M}}}^{(i)}\mathop  {\longrightarrow} \limits^{a.s.} {\log _2}\left( {1 + \frac{{\lambda \tau {E_\tau }{E_{\rm BS}}{{\left( {\beta _{{\rm{B - M}}}^{(i)}} \right)}^4}}}{{{{\sum\limits_{k = 1}^K {\left( {\beta _{{\rm{B - M}}}^{(k)}} \right)}^2}}}\sigma _0^2\left({\rm RD}_i+\sigma_0^2\right)}} \right)$, $R_{{\rm{0,S,M}}}^{(m,j)}\mathop  {\longrightarrow} \limits^{a.s.} {\log_2}\left( {1 + \frac{{\tau {E_\tau }{E_{\rm SC}}{{\left( {\beta _{{\rm{S - S}}}^{(m,m,j)}} \right)}^4}}}{{\sum\limits_{l = 1}^L {\left( {\beta _{{\rm{S - S}}}^{(m,m,l)}} \right)}^2 \sigma _0^2}\left({\rm RD}_{m,j}+\sigma_0^2\right)} }\right)$, $R_{{\rm{0,M,Z}}}^{(i)}\mathop  {\longrightarrow} \limits^{a.s.} {\log _2}\left( {1 + \frac{\lambda \tau {E_\tau }{E_{\rm BS}}/{\sum\limits_{k = 1}^K {\left( {\beta _{{\rm{B - M}}}^{(k)}} \right)}^{-2}}}{\sigma _0^2\left({\rm RD}_i+\sigma_0^2\right)}} \right)$ and $R_{{\rm{0,S,Z}}}^{(m,j)}\mathop  {\longrightarrow} \limits^{a.s.} {\log_2}\left( {1 + \frac{{\tau {E_\tau }}{E_{\rm SC}}/{\sum\limits_{l = 1}^L {\left( {\beta _{{\rm{S - S}}}^{(m,m,l)}} \right)}^{-2} }}{{\sigma _0^2\left({\rm RD}_{m,j}+\sigma_0^2\right)}} }\right)$ for MRT and ZFT, respectively, where the residual items are ${\rm RD}_i={E_{\rm BS}}{\beta _{{\rm{B - M}}}^{(i)}}+{E_{\rm SC}}\sum\limits_{n=1}^S{\beta _{{\rm{S - M}}}^{(n,i)}}$ and ${\rm RD}_{m,j}=E_{\rm SC}{\sum\limits_{n = 1}^M {\beta _{\rm{S - S}}^{(n,m,l)}}}+{E_{\rm BS}}\beta_{\rm B-S}^{(m,j)}$. It indicates that the channel estimation error induced interference, cross-tier and inter-SC interferences can not be eliminated when $p_\tau$ is scaled down proportionally to $\frac{1}{{N_{\rm SC}}}$ with fixed ${p_{\rm{BS}}}$ and ${p_{\rm{SC}}}$ in case II.

\emph{Proposition 3:} In case II as stated in Proposition 2, to achieve non-vanishing user rate as $N_{\rm SC} \to \infty$ with $N_{\rm BS} = \lambda N_{\rm SC}$ and the pilot reuse power $\gamma < S$, i.e., considering the pilot reuse introduced contamination, the pilot power scaling factor must satisfy $\theta = 0$. If $\theta>0$, the MUE rate $R_{{\rm{0,M,M}}}^{(i)}$ and $R_{{\rm{0,M,Z}}}^{(i)}$ still follow (\ref{sec3:hat_SCSI_MUE_rate_P2}) and (\ref{sec3:hat_SCSI_SUE_rate_P2}), while the asymptotic achievable rate expressions of the $j$th~($j\in J_m$) SUE in the $m$th~($m\in\{1,\dots,S\}$) SC for imperfect CSI based MRT and ZFT are
\begin{equation}\label{sec3:hat_SCSI_SUE_rate_P3}
\begin{split}
R_{{\rm{0,S,M}}}^{(m,j)}\mathop  {\longrightarrow} \limits_{{N_{\rm SC}} \to \infty }^{a.s.} 0,~R_{{\rm{0,S,Z}}}^{(m,j)}\mathop  {\longrightarrow} \limits_{{N_{\rm SC}} \to \infty }^{a.s.} 0.
\end{split}
\end{equation}

\section{User Scheduling Algorithms}\label{sec:US}
To maximize the sum rate of the scheduled MUEs and SUEs, exhaustive search in the whole MUE and SUE sets is one possible method to obtain optimal results. However, it is not practical since it has rather low searching speed with high complexity. In this section, $\gamma = S$, i.e., no pilot reuse, is assumed\footnote{For other pilot reuse factors, the corresponding user scheduling algorithm can be a similar way.}. As a traditional suboptimal method, a greedy scheduling algorithm is proposed according to the derived capacity lower bounds in Section~\ref{sec:Achiev_Rate}. Then, in comparison to the greedy scheduling algorithm, we propose a much simpler scheduling algorithm based on the obtained asymptotic results to maximize each cell's sum rate supposing that $N_{\rm SC}\to \infty$. It is called asymptotic scheduling algorithm which significantly reduces the computation complexity.

\subsection{Greedy Scheduling Algorithm (GSA)}
Aiming to obtain an optimal user scheduling algorithm, we formulate an optimization problem considering the maximization of total achievable rate for all the scheduled MUEs and SUEs, subject to the constraints on the scale of each UE set, i.e., (\ref{sec3:Opt1}),
\begin{table*}
\begin{subequations}
    \label{sec3:Opt1}
    \begin{align}
    \label{sec3:objective1}
    \begin{split}
    \mathop {\max }\limits_{I \subseteq {U_{{\rm{M}}}},{J_m} \subseteq U_{{\rm{S}}}^{(m)}} ~& {R_{\rm{SUM}}}\left( {I,{J_1}, \ldots ,{J_S}} \right) \buildrel \Delta \over = \frac{{T - \tau }}{T}\left( {\sum\limits_{i \in I} {R_{0,{\rm{M}}}^{(i)}}  + \sum\limits_{m = 1}^S {\sum\limits_{j \in {J_m}} {R_{0,{\rm{S}}}^{(m,j)}} } } \right)
    \end{split}\\
    \label{sec3:constraint11}
    \begin{split}
    {\rm{s.t.}}~~&\left| I \right| = K,~\left| {{J_m}} \right| = L,~m = 1, \ldots ,S
    \end{split}
    \end{align}
\end{subequations}
\end{table*}
which can surely be solved by inefficient exhaustive search. To reduce the computation complexity, a greedy scheduling algorithm is proposed, as summarized in Algorithm 1.  In one iteration, each cell schedules one additional user, the user that maximizes the total sum rate is added in one cell, given the scheduled users in all other cells.  This process repeats until the number of scheduled users in each cell reaches the target value.

As shown above, the proposed GSA is a suboptimal solution for the maximization of the sum rate, but it still results in high computation complexity, since each user's achievable rate is determined by the SCSI of the downlink channels from the BS and all SC nodes to the users, i.e., the BS and each SC node should share SCSI with one another even for each cell's own user scheduling.
\begin{table}[htbp]
\vspace{0em}
\label{sec3:tab:1}       
\centering
\begin{tabular}{p{0.45\textwidth}}
\toprule
\textbf{Algorithm 1:} Greedy scheduling algorithm \\
\midrule
\textbf{Initialization:} $N=\left\lfloor {\frac{K}{L}} \right\rfloor$, $I = \emptyset$, ${J_m} = \emptyset$, ${{\tilde U}_{{\rm{M}}}} = {U_{{\rm{M}}}}$ and $\tilde U_{{\rm{S}}}^{(m)} = U_{{\rm{S}}}^{(m)}$ for $m = 1,\dots,S$. \\
\textbf{Repeat:}\\
     For1 $k$ $=$ $1$ to $N$\\
     ~~~~~~${i^*} = \mathop {\arg \max }\limits_{i \in {{\tilde U}_{{\rm{M}}}}} {R_{{\rm{SUM}}}}\left( {I \cup \left\{ i \right\},{J_1}, \ldots ,{J_S}} \right)$\\
     ~~~~~~$I = I \cup \left\{ {{i^*}} \right\}$, ${{\tilde U}_{{\rm{M}}}} = {{\tilde U}_{{\rm{M}}}}\backslash \left\{ {{i^*}} \right\}$\\
     Endfor1\\
     For2 $m$ $=$ $1$ to $S$\\
     ~~~~~~$j_m^* = \mathop {\arg \max }\limits_{{j_m} \in \tilde U_{{\rm{S}}}^{(m)}} {R_{{\rm{SUM}}}}\left( {I,{J_1} \ldots ,{J_m} \cup \left\{ {{j_m}} \right\} \ldots ,{J_S}} \right)$\\
     ~~~~~~${J_m} = {J_m} \cup \left\{ {j_m^*} \right\}$, $\tilde U_{{\rm{S}}}^{(m)} = \tilde U_{{\rm{S}}}^{(m)}\backslash \left\{ {j_m^*} \right\}$\\
     Endfor2\\
\textbf{Until:} If $\left| {{J_m}} \right| = L$ with $m = 1, \ldots ,S$, change $N=K\bmod \left( L \right)$, go through For1 Loop and then stop;\\
\textbf{Output:} Output $I$ and ${J_m}$~($m=1,\dots,S$) as the solutions.\\
\bottomrule
\end{tabular}
\end{table}

\subsection{Asymptotic Scheduling Algorithm (ASA)}
In order to further reduce the computation complexity of the traditional GSA, a new algorithm named ASA is presented for MRT and ZFT, respectively, based on the asymptotic results as given in Subsection \ref{asymptotic}. From Propositions 1 and 2, it can be concluded that there is no inter-cell interference when $N_{SC}\to \infty$, i.e., each cell is able to do user scheduling according to its own statistical CSI and there is no information exchange requirement among the BS and SC nodes any more.

\subsubsection{Asymptotic Scheduling Algorithm for MRT (ASA-M)}
Since the Propositions 1 and 2 provide the asymptotic results for MRT, we define the achievable rate of the $i$th~($i\in U_{\rm{BS}}$) MUE and $j$th~($j\in U_{\rm{SC}}^{(m)}$) SUE in the $m$th~($m=1,\dots,S$) SC as $R_{{\rm 0,M,M-AS}}^{(i)}$ and $R_{{\rm{0,S,M-AS}}}^{(m,j)}$ according to (\ref{sec3:hat_SCSI_MUE_rate_P1}) or (\ref{sec3:hat_SCSI_MUE_rate_P2}) in Subsection \ref{asymptotic}. Since maximizing the sum rate is equivalent to maximizing each cell's rate, optimization problems for the MC and each SC are proposed on the constraints of the sizes for each selected user subset, respectively, i.e.,
\begin{subequations}
    \label{sec3:Opt21}
    \begin{align}
    \label{sec3:objective21}
    \begin{split}
    \mathop {\max }\limits_{I \subseteq {U_{{\rm{M}}}}}~ &{R_{{\rm{BS,M}}}}\left( I \right) \buildrel \Delta \over = \frac{{T - \tau }}{T}\sum\limits_{i \in I} {R_{0,{\rm{M,M-AS}}}^{(i)}}
    \end{split}\\
    \label{sec3:constraint21}
    \begin{split}
    {\rm{s.t.}}~~&\left| I \right| = K
    \end{split}
    \end{align}
\end{subequations}
\vspace{-1.5em}
\begin{subequations}
    \label{sec3:Opt22}
    \begin{align}
    \label{sec3:objective22}
    \begin{split}
    \mathop {\max }\limits_{{J_m} \subseteq U_{{\rm{S}}}^{(m)}} ~&R_{{\rm{SC,M}}}^{(m)}\left( J_m \right) \buildrel \Delta \over = \frac{{T - \tau }}{T}\sum\limits_{j \in {J_s}} {R_{0,{\rm{S,M-AS}}}^{(m,j)}}
    \end{split}\\
    \label{sec3:constraint22}
    \begin{split}
    {\rm{s.t.}}~~&\left| {{J_m}} \right| = L
    \end{split}
    \end{align}
\end{subequations}
which can be solved separately by exhaustive search. Similarly, as a suboptimal solution, a simplified greedy search method is proposed, in which the user scheduling at each cell is operated separately, i.e., it can be completed by its own node without sharing any CSI with other cells. The reader is referred to Algorithm 2 for a step-by-step summary of the proposed method.

\begin{table}[htbp]
\vspace{0em}
\label{sec3:tab:2}       
\centering
\begin{tabular}{p{0.45\textwidth}}
\toprule
\textbf{Algorithm 2:} Asymptotic scheduling algorithm for MRT\\
\midrule
\textbf{Initialization:} $I = \emptyset$, ${J_m} = \emptyset$, ${{\tilde U}_{{\rm{M}}}} = {U_{{\rm{M}}}}$ and $\tilde U_{{\rm{S}}}^{(m)} = U_{{\rm{S}}}^{(m)}$ for $m = 1,\dots,S$. \\
\textbf{Repeat1:}\\
     ~~~~~~${i^*} = \mathop {\arg \max }\limits_{i \in {{\tilde U}_{{\rm{M}}}}} {R_{{\rm{BS,M}}}}\left( I \cup \left\{ i \right\}\right)$\\
     ~~~~~~$I = I \cup \left\{ {{i^*}} \right\}$, ${{\tilde U}_{{\rm{M}}}} = {{\tilde U}_{{\rm{M}}}}\backslash \left\{ {{i^*}} \right\}$\\
\textbf{Until1:} Stop if $\left| {{I}} \right| = K$.\\
For $m$ $=$ $1$ to $S$\\
\textbf{Repeat2:}\\
     ~~~~~~$j_m^* = \mathop {\arg \max }\limits_{{j_m} \in \tilde U_{{\rm{S}}}^{(m)}} R_{{\rm{SC,M}}}^{(m)}\left( {J_m}\cup \left\{ {{j_m}} \right\} \right)$\\
     ~~~~~~${J_m} = {J_m} \cup \left\{ {j_m^*} \right\}$, $\tilde U_{{\rm{S}}}^{(m)} = \tilde U_{{\rm{S}}}^{(m)}\backslash \left\{ {j_m^*} \right\}$\\
\textbf{Until2:} Stop if $\left| {{J_m}} \right| = L$.\\
Endfor\\
\textbf{Output:} Output $I$ and ${J_m}$~($m=1,\dots,S$) as the solutions.\\
\bottomrule
\end{tabular}
\end{table}

\subsubsection{Asymptotic Scheduling Algorithm for ZFT (ASA-Z)}
Likewise, since the Propositions 1 and 2 also provide the asymptotic results for ZFT, we can define the achievable rate of the $i$th~($i\in U_{\rm{BS}}$) MUE and $j$th~($j\in U_{\rm{SC}}^{(m)}$) SUE in the $m$th~($m=1,\dots,S$) SC as $R_{{\rm{0,M,Z-AS}}}^{(i)}$ and $R_{{\rm{0,S,Z-AS}}}^{(m,j)}$ according to (\ref{sec3:hat_SCSI_SUE_rate_P1}) or (\ref{sec3:hat_SCSI_SUE_rate_P2}) in Subsection \ref{asymptotic}. However, notably, $R_{{\rm{0,M,Z-AS}}}^{(1)}=\cdots = R_{{\rm{0,M,Z-AS}}}^{(K)}$ for fixed ${\sum\limits_{k = 1}^K {\left( {\beta _{{\rm{B - M}}}^{(k)}} \right)}^{-\eta}}$, and $R_{{\rm{0,S,Z-AS}}}^{(m,1)}=\cdots =R_{{\rm{0,S,Z-AS}}}^{(m,L)}$ for fixed ${{\sum\limits_{l = 1}^L {\left( {\beta _{{\rm{S - S}}}^{(m,m,l)}} \right)}^{-\eta} }}$ with $m=1,\cdots,S$\footnote{Here, $\eta = 1$ for case I in (\ref{sec3:hat_SCSI_SUE_rate_P1}) and $\eta = 2$ for case II in (\ref{sec3:hat_SCSI_SUE_rate_P2}).}, which means that maximizing the sum rate is equivalent to
\begin{subequations}
    \label{sec3:Opt31}
    \begin{align}
    \label{sec3:objective31}
    \begin{split}
    \mathop {\min }\limits_{I \subseteq {U_{{\rm{M}}}}}~ &{\sum\limits_{k\in I} {\left( {\beta _{{\rm{B - M}}}^{(k)}} \right)}^{-\eta}}
    \end{split}\\
    \label{sec3:constraint31}
    \begin{split}
    {\rm{s.t.}}~~&\left| I \right| = K
    \end{split}
    \end{align}
\end{subequations}
\vspace{-1.5em}
\begin{subequations}
    \label{sec3:Opt32}
    \begin{align}
    \label{sec3:objective32}
    \begin{split}
    \mathop {\min }\limits_{{J_m} \subseteq U_{{\rm{S}}}^{(m)}} ~&{{\sum\limits_{l \in J_m} {\left( {\beta _{{\rm{S - S}}}^{(m,m,l)}} \right)}^{-\eta} }}
    \end{split}\\
    \label{sec3:constraint32}
    \begin{split}
    {\rm{s.t.}}~~&\left| {{J_m}} \right| = L.
    \end{split}
    \end{align}
\end{subequations}
Different from ASA-M, the optimal ASA-Z can be realized by the max-beta based scheduling algorithm, the process of which is similar to the Algorithm 2. To solve this optimization problems of (\ref{sec3:Opt31}) and (\ref{sec3:Opt32}), we just need to find $K$ MUEs with the first $K$-maximum ${\beta _{{\rm{B - M}}}^{(k)}}$ and $L$ SUEs with the first $L$-maximum ${\beta _{{\rm{S - S}}}^{(m,m,l)}}$, respectively. Hence, we replace the item ${R_{{\rm{BS,M}}}}\left( I \cup \left\{ i \right\}\right)$ and $R_{{\rm{SC,M}}}^{(m)}\left( {J_m}\cup \left\{ {{j_m}} \right\} \right)$ in the Algorithm 2 by ${\beta _{{\rm{B - M}}}^{(i)}}$ and ${\beta _{{\rm{S - S}}}^{(m,m,j_m)}}$, respectively, to realize the optimal ASA-Z, which is much simpler than ASA-M as neither achievable rate calculation nor SCSI exchange is needed at each node.

\section{Numerical Results}\label{sec:NumericalResults}
In this section, we present numerical results to validate the derived achievable rate expressions and examine the performance of HetNet with large-scale antenna arrays in downlink channels. A dense tier of $S$ SCs that are uniformly distributed in a circle as shown in Fig.~\ref{Section3_1}. Suppose that the users in the cell are uniformly distributed with the total number $200$ and $500$ for $S=8$ and $S=20$, respectively, and that the MUEs and SUEs are identified by the bias user association. The number of scheduled MUEs and SUEs in each SC are $K=20$ and $L=4$, respectively, with training length $\tau=K+L\times \gamma$ and bias factors $\kappa_{\rm BS} = 1$ and $\kappa_{\rm SC} = 1.2$~\cite{Lin2015} for BS and SC nodes, respectively. Also we assume that equal transmit power at the each SC node satisfies $p_{\text{SC}}^{(1)}$~(dBm)$=\dots=p_{\text{SC}}^{(S)}$~(dBm)$=p_{\text{BS}}$~(dBm)$-22$~(dB) as $p_{\text{BS}}$ changes, and that $N_{\text{BS}}/N_{\text{SC}} = \lambda = 10$ is fixed as $N_{\text{BS}}$ changes. See Table~\ref{sec3:tab:4} for simulation parameters and assumption details.
\begin{table}[htbp]
\caption{\textsc{Simulation Parameters}}
\vspace{0em}
\label{sec3:tab:4}       
\centering
\begin{tabular}[t]{p{105pt}p{125pt}}
\hline\noalign{\smallskip}
Parameters & Setting  \\
\noalign{\smallskip}\hline\noalign{\smallskip}
    Bandwidth & 20~MHz \\
    Macro Cell radius & 1000~m\\
    Distance (BS,SCs) & 800~m\\
    The number of SCs & $S=8~{\rm or}~20$ \\
    Pilot Reuse factor & $\gamma=1\sim8~{\rm or}~1\sim20$ \\
    Coherent interval & $T=200$ \\
    Transmit power threshold (BS/SC) & ${p_{{\rm{BS}}}}-{p_{{\rm{SC}}}}=22$~dB, ${p_{{\rm{SC}}}^{(1)}}=\cdots={p_{{\rm{SC}}}^{(S)}}={p_{{\rm{SC}}}}$ \\
    Noise power density & -174~dBm/Hz  \\
    Number of MUEs/SUEs & $K=20$, $L=4$\\
    Pathloss model (BS) & $\theta_{\rm{BS}}(d)=128.1+37.6\text{log}_{10}(d)$, $d$~(km)~\cite{3GPP2011}\\
    Pathloss model (SC) & $\theta_{\rm{SC}}(d)=140.7+36.7\text{log}_{10}(d)$, $d$~(km)~\cite{3GPP2011}\\
  \hline
\end{tabular}
\end{table}
\vspace{-1em}

\subsection{Comparison between One-Tier and Two-Tier Network Topologies}
\begin{figure}
   \centering
   \includegraphics[scale=0.5]{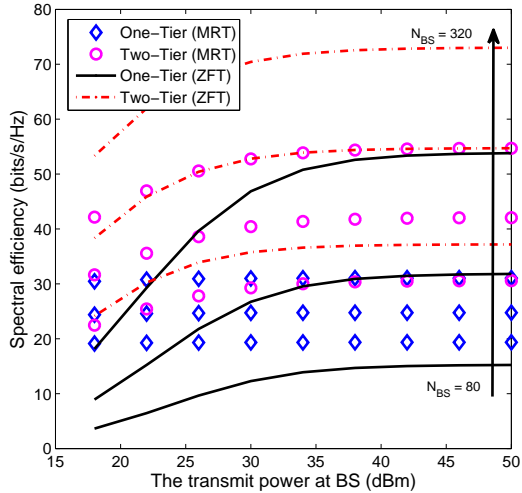}
   \caption{Spectral efficiency versus $p_{\text{BS}}$ for one-tier and two-tier network topologies (RSA, $p_{\tau}=0$~dBm, $S=8$, $\gamma=8$).}
   \label{Section3_21}
\end{figure}

\begin{figure}
   \centering
   \includegraphics[scale=0.5]{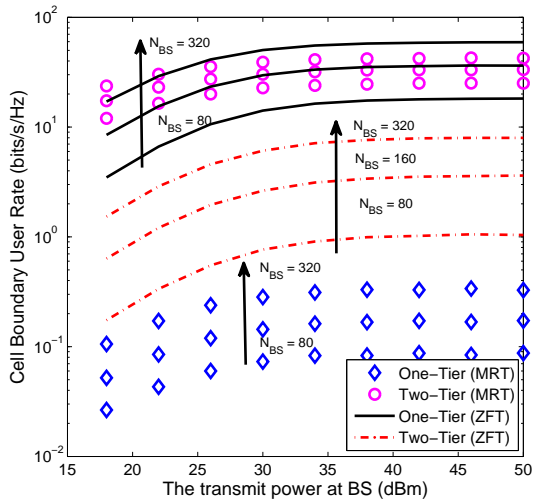}
   \caption{Cell boundary SC user rate versus $p_{\text{BS}}$ for one-tier and two-tier network topologies (RSA, $p_{\tau}=0$~dBm, $S=8$, $\gamma=8$).}
   \label{Section3_22}
\end{figure}

First, the effectiveness of the considered large-scale two-tier model is demonstrated by comparing it with the one-tier network topology. Here, suppose that simple random scheduling algorithm (RSA) is utilized for user scheduling, $p_{\tau} = 0$ dBm, $S=8$ and $\gamma=8$. To be fair, in the one-tier network the number of antennas is assumed to satisfy $N = N_{\text{BS}} + S\times N_{\text{SC}}$ and the number of UEs is set to be $K+S\times L$ including both MUEs and SUEs in the two-tier network. Fig.~\ref{Section3_21} shows that the two-tier network outperforms one-tier network on the spectral efficiency with both MRT and ZFT precoding schemes in the downlink channels. As an indication of coverage, Fig.~\ref{Section3_22} plots the cell boundary small cell user rate versus $p_{\text{BS}}$ with fixed $N_{\text{BS}}=80$, $160$ and $320$, where the cell boundary users are composed of all the SUEs identified by the biased user association~\cite{Lin2015}. For MRT the two-tier network has much higher cell boundary user rate, indicating better coverage than one-tier. However, for ZFT the two-tier network underperforms the one-tier for cell boundary SUEs, since neither the inter-SC interference nor the cross-tier interference can be cancelled in the non-cooperative two-tier HetNet systems. While in the one-tier network, the inter-user interferences could be totally eliminated by the ZFT precoding without inter-SC and cross-tier interferences. On the other hand, the ZFT precoding of the one-tier network is much more complex than that of the two-tier network due to the larger number of UEs ($K+S\times L$) served simultaneously by the BS.

\subsection{Validation of Lower Capacity Bounds and Pilot Reuse Pattern}
In this subsection, the effectiveness of the derived capacity lower bounds with imperfect CSI based MRT in (\ref{sec3:hat_SCSI_MUE_rate2_CF}) and (\ref{sec3:hat_SCSI_SUE_rate2_CF}) and ZFT in (\ref{sec3:hat_SCSI_MUE_rate2_CF2}) and (\ref{sec3:hat_SCSI_SUE_rate2_CF2}) is evaluated by comparing them with the Monte-Carlo simulation results. Then, the performance of the presented simple PR pattern is evaluated. Besides, the asymptotic analyses with massive arrays for the two cases in Propositions 1 and 2 are examined in this subsection. Here, random scheduling algorithm (RSA) is utilized.

\begin{figure}
   \centering
   \includegraphics[scale=0.5]{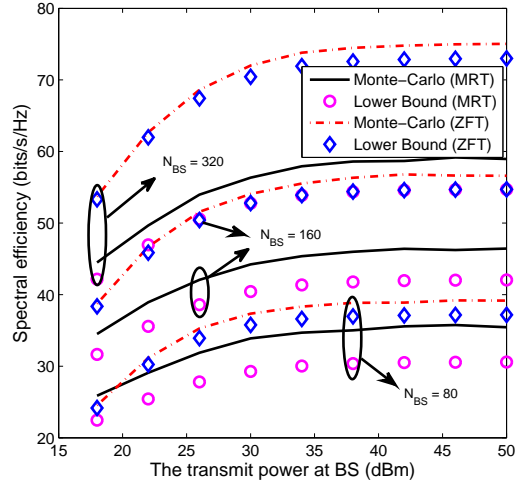}
   \caption{Spectral efficiency versus $p_{\text{BS}}$ for Monte-Carlo results and lower bounds (RSA, $p_{\tau}=0$~dBm, $S=8$, $\gamma=8$).}
   \label{Section3_2}
\end{figure}

Firstly, $p_{\tau} = 0$ dBm, $S=8$ and $\gamma=8$ are set. In Fig.~\ref{Section3_2}, the spectral efficiency versus $p_{\text{BS}}$ curves with fixed $N_{\text{BS}}=80$, $160$ and $320$ for capacity lower bounds are compared with those obtained from (\ref{sec3:hat_ICSI_MUE_rate}) and (\ref{sec3:hat_ICSI_SUE_rate}) by Monte-Carlo simulation. ``Lower Bound (MRT)'' indicates the capacity lower bound obtained by (\ref{sec3:hat_SCSI_MUE_rate2_CF}) and (\ref{sec3:hat_SCSI_SUE_rate2_CF}), and ``Lower Bound (ZFT)'' is calculated from (\ref{sec3:hat_SCSI_MUE_rate2_CF2}) and (\ref{sec3:hat_SCSI_SUE_rate2_CF2}). It can be observed from Fig.~\ref{Section3_2} that the relative performance gap between ``Lower Bound (ZFT)'' and ``Monte-Carlo (ZFT)'' is quite smaller than that between ``Lower Bound (MRT)'' and ``Monte-Carlo (MRT)'', especially at lower transmit power with larger number of transmit antennas. Moreover, the spectral efficiency of ZFT increases much faster than that of MRT as $p_{\text{BS}}$ increases, due to the fact that the effect of interference is much larger than that of the noise for higher SNR while ZFT is able to null multi-user interference signals~\cite{Hong2013}. In this way, the derived capacity lower bounds by Jensen's inequality are proved to be accurate predictors of the system performance.

\begin{figure}
   \centering
   \includegraphics[scale=0.5]{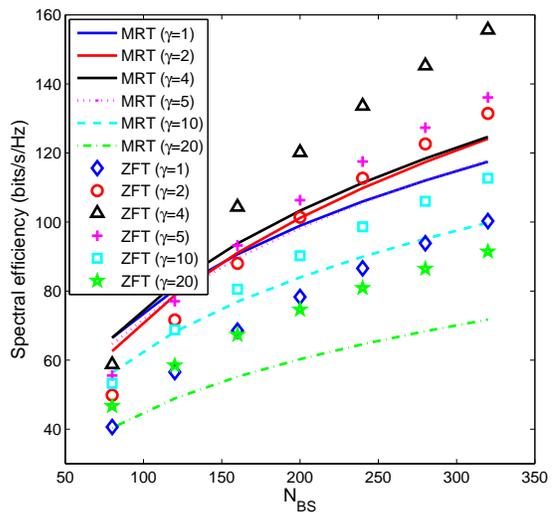}
   \caption{Spectral efficiency versus $N_{\text{BS}}$ for different PR factors (RSA, $p_{\rm BS}=46$~dBm, $p_{\tau}=0$~dBm, $S=20$).}
   \label{Section3_5}
\end{figure}

Secondly, setting $p_{\rm BS}=46$ dBm, $p_{\tau}=0$ dBm and $S=20$, the spectral efficiency of the presented PR pattern with different pilot reuse factor $\gamma$ versus the number of transmit antennas at BS is illustrated in Fig.~\ref{Section3_5}. When the spectral efficiency is calculated by (\ref{sec3:hat_SCSI_MUE_rate2_CF}) and (\ref{sec3:hat_SCSI_SUE_rate2_CF}) taking the influence of the pilot overhead $\tau$ into consideration, Fig.~\ref{Section3_5} indicates that PR factor $\gamma=4$ yields the best performance for both MRT and ZFT, which achieves the optimal trade off between pilot overhead and pilot contamination from the reuse of the pilot sets. Notably, when the PR factor is reduced to $\gamma = 2$ and $\gamma = 1$, the spectral efficiency of ZFT is no longer larger than that of MRT as shown in Fig.~\ref{Section3_5}, which indicates that pilot reuse introduced pilot contamination effect is more severe for ZFT than that for MRT. It can be explained by the fact that more pilot reuse increases the channel estimation error, and thus monotonically decreases the SINR of the SUEs in the ZFT case as stated by Remark 2.

\begin{figure}
   \centering
   \includegraphics[scale=0.5]{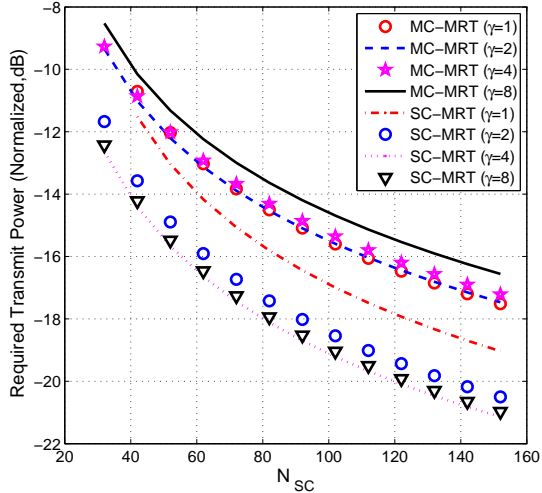}
   \caption{Transmit power required to achieve 1 bit/s/Hz per user for case I (normalized $p_{\tau}=0$~dB, $S=8$).}
   \label{Section3_3_1}
\end{figure}

\begin{figure}
   \centering
   \includegraphics[scale=0.5]{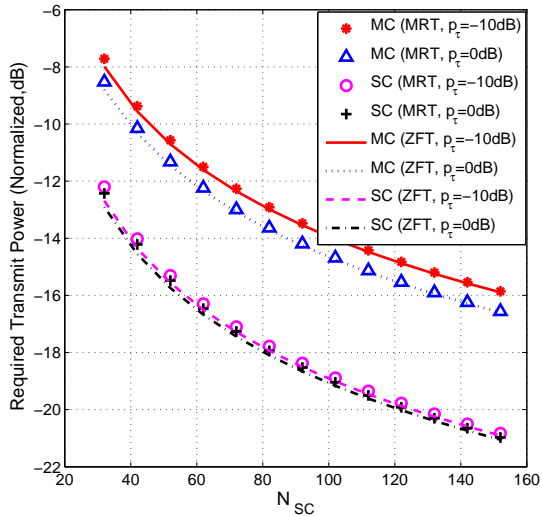}
   \caption{Transmit power required to achieve 1 bit/s/Hz per user for case I ($S=8$, $\gamma=8$).}
   \label{Section3_3_2}
\end{figure}

\begin{figure}
   \centering
   \includegraphics[scale=0.5]{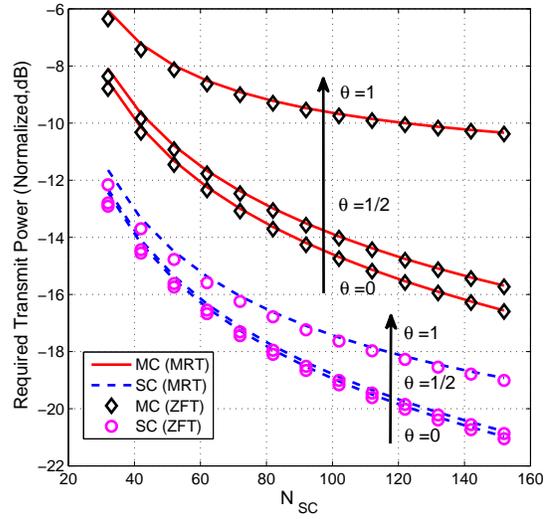}
   \caption{Transmit power required to achieve 1 bit/s/Hz per user for case II (normalized $E_{\tau}=0$~dB, $S=8$, $\gamma=8$).}
   \label{Section3_3_3}
\end{figure}

\begin{figure}
   \centering
   \includegraphics[scale=0.5]{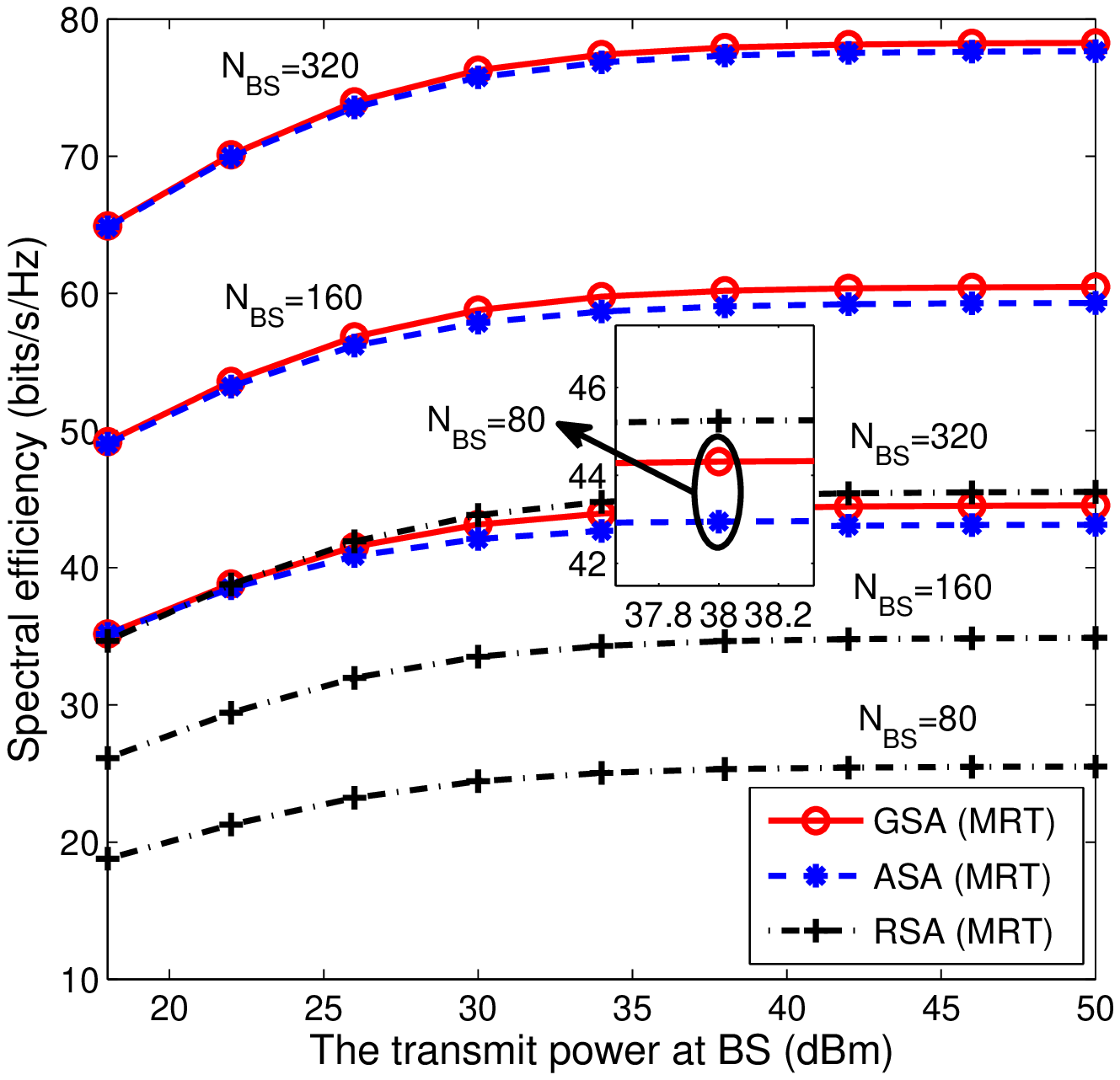}
   \caption{Spectral efficiency versus $p_{\text{BS}}$ for different user scheduling algorithms (MRT, $p_{\tau}=5$~dBm, $S=8$, $\gamma=8$).}
   \label{Section3_6}
\end{figure}

\begin{figure}
   \centering
   \includegraphics[scale=0.5]{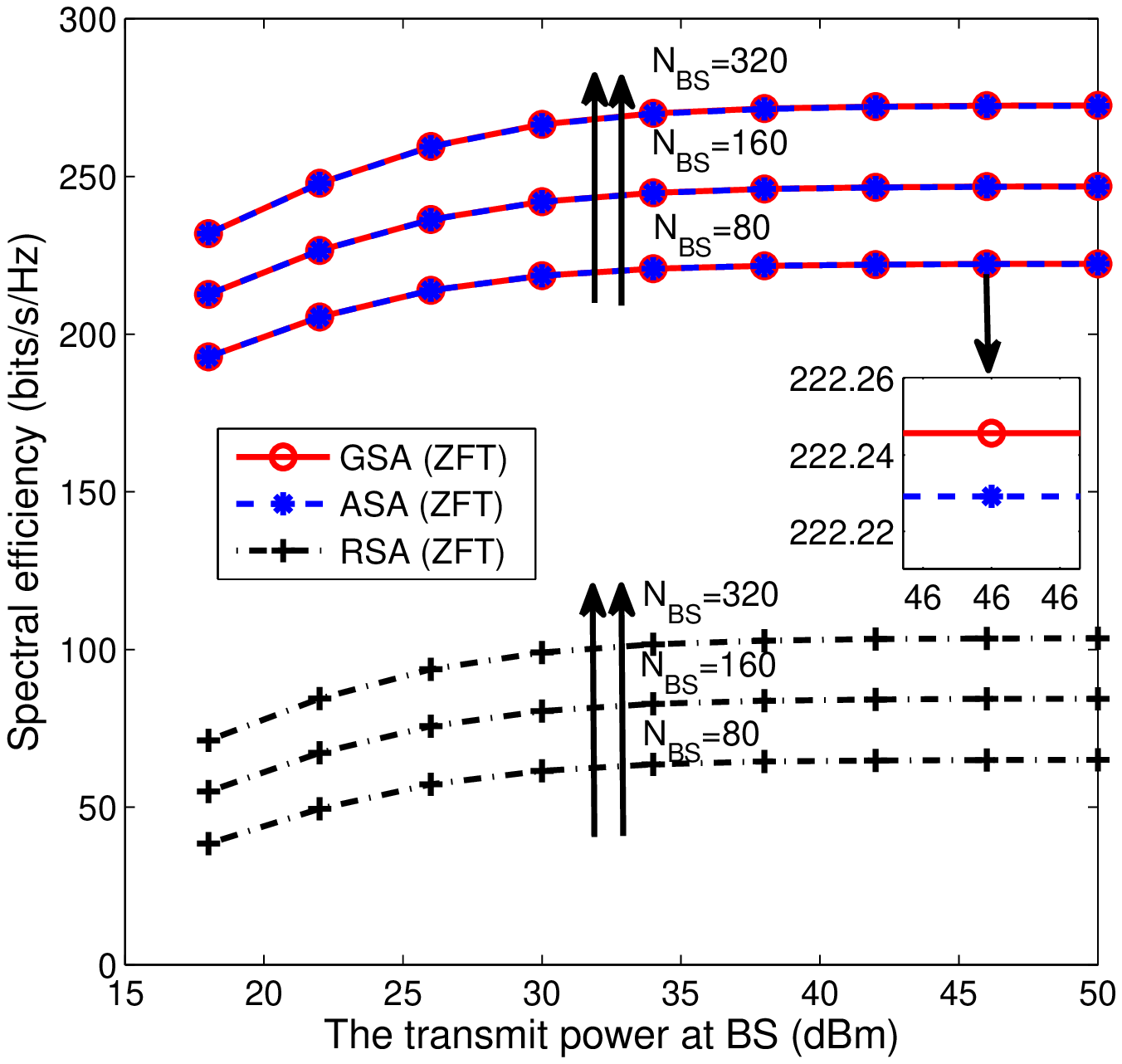}
   \caption{Spectral efficiency versus $p_{\text{BS}}$ for different user scheduling algorithms (ZFT, $p_{\tau}=5$~dBm, $S=8$, $\gamma=8$).}
   \label{Section3_7}
\end{figure}

Finally, the asymptotic analyses with massive arrays for the two cases in Propositions 1 and 2 are examined with the MUE and SUE large-scale fading factors fixed as $\beta_{\rm B-M} = 1$, $\beta_{\rm B-S} = 0.2$, $\beta_{\rm S-S}^{(m,m)} = 5$, $\beta_{\rm S-S}^{(m,n)} = 0.6$~($m\ne n$) and $\beta_{\rm S-M} = 0.6$ considering pathloss, noise variance $\sigma_0^2 = 1$ and the normalized pilot power $p_{\tau}=0$~dB. Fig.~\ref{Section3_3_1} shows the required MC and SC transmit power $p_{\rm BS}$ and $p_{\rm SC}$ to achieve $1$~bit/s/Hz per MUE and SUE, respectively, for MRT in case I. It is obvious from Fig.~\ref{Section3_3_1} that in case I where the pilot power $p_{\tau}$ is fixed, the required $p_{\rm BS}$ and $p_{\rm SC}$ are significantly reduced as $N_{\rm SC}$ increases, and that the required $p_{\rm BS}$ with $\gamma = 1$ is the lowest and the $p_{\rm SC}$ with $\gamma = 4$ is the lowest in comparison to other PR factors. Moreover, for $\gamma < 4$ in Fig.~\ref{Section3_3_1}, it is evident that the lower the PR factor $\gamma$ is, the higher $p_{\rm SC}$ is required to achieve $1$~bit/s/Hz per user. The observation indicates that the pilot contamination effect existed at the SCs increases the required transmit power $p_{\rm SC}$ to achieve $1$~bit/s/Hz per SUE. Regarding the imperfect CSI effect, Fig.~\ref{Section3_3_2} shows that less transmit power in both MC and SCs are required when $p_{\tau}$ is high. For case II with $E_{\tau}=0$~dB and the pilot power scaling down by ${p_{\tau}} = \frac{{{E_{\tau}}}}{N_{\rm SC}^\theta}$, Fig.~\ref{Section3_3_3} shows that higher $\theta$ leads to more slowly reduced $P_{\tau}$, because the imperfect CSI effect becomes more severe when the pilot power is reduced much faster with the increase of $N_{\rm SC}$.

\subsection{User Scheduling}
In this subsection, the proposed user scheduling algorithms are examined with respect to the spectral efficiency of the HetNet downlink systems. Here, we choose $p_\tau=5$ dBm, $S=8$ and $\gamma=8$ for comparison fairness.

First, simulations are performed for MRT on the proposed two user scheduling algorithms in regard to the spectral efficiency versus $p_{\text{BS}}$, which is given in Fig.~\ref{Section3_6} with fixed $N_{\text{BS}}=80$, $160$ and $320$. We obtain the results that GSA outperforms ASA and RSA even when $N_{\text{BS}}$ is not sufficiently high, and that the curve of ASA approaches to the curve of GSA as $N_{\text{BS}}$ increases. Furthermore, the performance gap between ASA and RSA for MRT is large. Fig.~\ref{Section3_7} shows the spectral efficiency versus $p_{\text{BS}}$ for ZFT on two different user scheduling schemes. It can be observed that the two curves for GSA and ASA almost coincide with each other even when the number of antennas is not very large, which demonstrates the effectiveness of ASA. Hence, for user scheduling of both MRT and ZFT, ASA is a good choice that achieves better performance with lower complexity. Moreover, by comparing Fig.~\ref{Section3_2} with Fig.~\ref{Section3_6} and Fig.~\ref{Section3_7}, it is shown that increasing $p_\tau$ from 0 dBm to 5 dBm introduces better performance for ZFT than MRT, indicating that different imperfect CSI effect on ZFT and MRT.

\section{Conclusion}\label{sec:conclusion}
In this paper, we have investigated the performance of a two-tier network with large-scale antenna arrays set at both BS and SCs. With MRT precoding employed at each node, we have derived capacity lower bounds with closed-form expressions for both imperfect CSI based MRT and ZFT cases, where a simple pilot reuse pattern is utilized for the channel estimation procedure to obtain the estimated imperfect CSI, followed by asymptotic analyses. The benefits of employing large number of antennas at both BS and SCs have been demonstrated. Simulation results have shown that the derived closed-form expressions for the achievable rate are accurate predictors of the system performance for both MRT and ZFT, and that more pilot reuse is able to yield higher effective performance under different system configurations even though pilot contamination exists. As for user scheduling, two schemes have been proposed. The greedy scheduling algorithm designed based on the derived capacity lower bounds only requires the statistical CSI but not the instantaneous CSI exchange among BS and SCs. The asymptotic scheduling algorithm, based on the asymptotic analysis results, has even lower complexity by removing the need for any CSI exchange among nodes, and can still achieve near-optimal performance for both MRT and ZFT in the asymptotic regime of massive antenna arrays. Furthermore, it has been found that in the asymptotic regime and when ZFT is used, the capacity lower bound of a user is proportional to the large-scale fading factor of the channel from the user to its base station. Therefore, ASA-Z can further reduce complexity by circumventing the need to calculate the achievable rate.

\section{Proof of (\ref{sec3:hat_alpha2})}\label{sec3:Appendix_A}
As to the derivation of (\ref{sec3:hat_alpha2}), we use the identity~\cite{Timm2002,Graczyk2003}
\begin{equation}\label{sec3:inv_Wishart_definition}
{\rm {E}}\left[  {\bf{W}}^{ - 1} \right] = \frac{{{{\bf{\Sigma }}^{ - 1}}}}{{n - m - 1}}
\end{equation}
where ${\bf{W}} \sim \mathcal{W}_m({\bf{\Sigma }},n)$ is an $m\times m$ central complex Wishart matrix with $n$~($n> m$) degrees of freedom and the distribution of ${\bf{W}}^{ - 1}$ is called an inverted Wishart distribution, following $\mathcal{W}_m^{-1}({\bf{\Sigma }}^{-1},n)$. It can be easily concluded that ${{\left( {{{\hat{\bf G}}_{\rm B-M}^H}\hat{\bf{G}}_{\rm B-M}} \right)}^{ - 1}}\sim \mathcal{W}_{K}^{-1}({\hat{\bf{D }}}_{\rm B-M}^{-1},N_{\rm BS})$ and ${{\left[ {{\left({\hat{\bf G}}_{\rm S-S}^{(m,m)}\right)^H}\hat{\bf{G}}_{\rm S-S}^{(m,m)}} \right]}^{ - 1}}\sim \mathcal{W}_{L}^{-1}(\left({\hat{\bf{D }}}_{\rm S-S}^{(m,m)}\right)^{-1},N_{\rm SC})$ for $\forall m\in\{1,\cdots,S\}$, hence
\begin{equation}\label{sec3:inv_Wishart}
\begin{split}
&{\rm {E}}\left[ {{\left( {{{\hat{\bf G}}_{\rm B-M}^H}\hat{\bf{G}}_{\rm B-M}} \right)}^{ - 1}} \right] = \frac{{{{\hat{\bf{D}}}_{\rm B-M}^{ - 1}}}}{{N_{\rm BS} - K - 1}}\\
&{\rm {E}}\left[ {{\left[ {{\left({\hat{\bf G}}_{\rm S-S}^{(m,m)}\right)^H}\hat{\bf{G}}_{\rm S-S}^{(m,m)}} \right]}^{ - 1}} \right] = \frac{\left({\hat{\bf{D }}}_{\rm S-S}^{(m,m)}\right)^{-1}}{{N_{\rm SC} - L - 1}}
\end{split}
\end{equation}
where $N_{\rm BS}>K$ and $N_{\rm SC}>L$. In this way, we have
\begin{equation}\label{sec3:inv_Wishart_alpha}
\begin{split}
&{{\alpha} _{{\rm{BS}}}} = \sqrt {\frac{\left( {{N_{{\rm{BS}}}} - K - 1} \right)P_{\rm BS}}{{\Psi }_{\rm{B - M}}}}, {\alpha} _{{\rm{SC}}}^{(m)} = \sqrt {\frac{{\left( {{N_{{\rm{SC}}}} - L - 1} \right)P_{{\rm{SC}}}^{(m)}}}{{\Psi _{{\rm{S - S}}}^{(m)}}}}
\end{split}
\end{equation}
where ${{\Psi }_{{\rm{B - M}}}}{\rm{ = }}\sum\limits_{i = 1}^K {\frac{1}{{\hat \beta _{{\rm{B - M}}}^{(i)}}}}$, and $\Psi _{{\rm{S - S}}}^{(m)} = \sum\limits_{l = 1}^L {\frac{1}{{\hat \beta _{{\rm{S - S}}}^{(m,m,l)}}}}$ with $m \in\left\{1,\dots,S\right\}$.

\section{Proof of Theorem 1}\label{sec3:Appendix_B}
To derive the closed-form expression of the MUE achievable rate in (\ref{sec3:hat_SCSI_MUE_rate2_CF}) based on Jensen's inequality, we start from the expectation of the SINR's reciprocal, given by (\ref{sec3:mean_inverse_MUE}), where the item of $\frac{{{\rm {E}}\left[ {{\rm{CT}}{{\rm{I}}_i}} \right] + \sigma _0^2}}{{\alpha _{{\rm{BS}}}^2}}$ does not depend on ${{\hat { \bf g}}_{{\rm{B - M}}}^{(i)}}$. Since large-scale antenna arrays are set at both the BS and SCs, some results from Gaussian distributed estimated channel in (\ref{sec3:estimated_CSI_exp})\footnote{Due to the estimated channel model in (\ref{sec3:estimated_CSI_exp}), we have that $\hat{\bf g}_i$ and $\hat{\bf g}_j$ are mutually independent $N\times 1$ vectors with $\forall i\ne j$ whose elements are i.i.d. zero-mean Gaussian distributed with variances $\hat\sigma_i^2$ and $\hat\sigma_j^2$, respectively. Then, it can be concluded that ${\rm E}\left[{\hat{\bf{g}}_i^H}\hat{\bf{g}}_i\right]=N\hat\sigma _i^2$, ${\rm E}\left[{\hat{\bf{g}}_j^H}\hat{\bf{g}}_j\right]=N\hat\sigma _j^2$, and ${\rm E}\left[{\hat{\bf{g}}_i^H}\hat{\bf{g}}_j\right]=0$. Also, we can obtain that ${\rm E}\left[|{\hat{\bf{g}}_i^H}\hat{\bf{g}}_j|^2\right]=N\hat\sigma _i^2\hat\sigma _j^2$.}~\cite{Cramer1970}
can be utilized. Using \cite[Lemma~2.9]{Tulino2004} (\ref{sec3:Lemma2_9}), where ${\bf{W}} \sim {\mathcal{W}_m}\left( {{{\bf{I}}_n},n} \right)$ is an $m\times m$ central complex Wishart matrix with $n$~($n>m+1$) degrees of freedom, we have (\ref{sec3:mean_inv_24_MUE}).

\begin{table*}
\begin{equation}\label{sec3:mean_inverse_MUE}
\begin{split}
    &{\rm {E}}\left[ {\frac{{{\rm{EE}}{{\rm{I}}_i} + {\rm{IM}}{{\rm{I}}_i} + {\rm{CT}}{{\rm{I}}_i} + \sigma _0^2}}{{\alpha _{{\rm{BS}}}^2{{\left\| {{\hat { \bf g}}_{{\rm{B - M}}}^{(i)}} \right\|}^4}}}} \right] = {\rm {E}}\left[ {\frac{{{\rm{EE}}{{\rm{I}}_i}}}{{\alpha _{{\rm{BS}}}^2{{\left\| {{\hat { \bf g}}_{{\rm{B - M}}}^{(i)}} \right\|}^4}}}} \right] + {\rm {E}}\left[ {\frac{{{\rm{IM}}{{\rm{I}}_i}}}{{\alpha _{{\rm{BS}}}^2{{\left\| {{\hat { \bf g}}_{{\rm{B - M}}}^{(i)}} \right\|}^4}}}} \right] + {\rm {E}}\left[ {\frac{1}{{{{\left\| {{\hat { \bf g}}_{{\rm{B - M}}}^{(i)}} \right\|}^4}}}} \right]\frac{{{\rm {E}}\left[ {{\rm{CT}}{{\rm{I}}_i}} \right] + \sigma _0^2}}{{\alpha _{{\rm{BS}}}^2}}
\end{split}
\end{equation}
\begin{equation}\label{sec3:Lemma2_9}
\begin{split}
    {\rm {E}}\left[ {{\rm {Tr}}\left\{ {{{\bf{W}}^{ - 1}}} \right\}} \right] &= \frac{m}{{n - m}},~{\rm {E}}\left[ {T{r^2}\left\{ {{{\bf{W}}^{ - 1}}} \right\}} \right] = \frac{m}{{n - m}}\left( {\frac{n}{{{{\left( {n - m} \right)}^2} - 1}} + \frac{{m - 1}}{{n - m + 1}}} \right)
\end{split}
\end{equation}
\begin{equation}\label{sec3:mean_inv_24_MUE}
\begin{split}
{\rm {E}}\left[ {\frac{1}{{{{\left\| {{\hat { \bf g}}_{{\rm{B - M}}}^{(i)}} \right\|}^2}}}} \right] =\frac{1}{{({N_{{\rm{BS}}}} - 1)\hat\beta _{{\rm{B - M}}}^{(i)}}},~{\rm {E}}\left[ {\frac{1}{{{{\left\| {{\hat { \bf g}}_{{\rm{B - M}}}^{(i)}} \right\|}^4}}}} \right] =\frac{1}{{({N_{{\rm{BS}}}} - 1)({N_{{\rm{BS}}}} - 2){{\left( {\hat\beta _{{\rm{B - M}}}^{(i)}} \right)}^2}}}.
\end{split}
\end{equation}
\begin{equation}\label{sec3:hat_mean_inv_PNI_MUE}
\begin{split}
{\rm {E}}\left[ {\frac{{{\rm{EE}}{{\rm{I}}_i}}}{{\alpha _{{\rm{BS}}}^2{{\left\| {{\hat { \bf g}}_{{\rm{B - M}}}^{(i)}} \right\|}^4}}}} \right] = {\rm {E}}\left[ {\frac{1}{{{{\left\| {{\hat { \bf g}}_{{\rm{B - M}}}^{(i)}} \right\|}^2}}}} \right]{\rm {E}}\left[ {{{\left| {\tilde \xi _{{\rm{B - M}}}^{(i)}} \right|}^2}} \right] = \frac{{\beta _{{\rm{B - M}}}^{(i)} - \hat \beta _{{\rm{B - M}}}^{(i)}}}{{({N_{{\rm{BS}}}} - 1)\hat \beta _{{\rm{B - M}}}^{(i)}}}
\end{split}
\end{equation}
\begin{equation}\label{sec3:hat_mean_inv_IMI_MUE}
\begin{split}
&{\rm {E}}\left[ {\frac{{{\rm{IM}}{{\rm{I}}_i}}}{{\alpha _{{\rm{BS}}}^2{{\left\| {{\hat { \bf g}}_{{\rm{B - M}}}^{(i)}} \right\|}^4}}}} \right]= {\rm {E}}\left[ {\frac{1}{{{{\left\| {{\hat { \bf g}}_{{\rm{B - M}}}^{(i)}} \right\|}^4}}}} \right]\sum\limits_{k \ne i}^K {{\rm {E}}\left[ {{{\left| {{{\left( {{\bf{\xi }}_{{\rm{B - M}}}^{(i)}} \right)}^T}{{\left( {{\hat { \bf g}}_{{\rm{B - M}}}^{(k)}} \right)}^*}} \right|}^2}} \right]}\\
&+{\rm {E}}\left[ {\frac{1}{{{{\left\| {{\hat { \bf g}}_{{\rm{B - M}}}^{(i)}} \right\|}^2}}}} \right]\sum\limits_{k \ne i}^K {{\rm {E}}\left[ {{{\left| {{\hat {\tilde g}}_{{\rm{B - M}}}^{(k)}} \right|}^2}} \right]}= \frac{{\sum\limits_{k \ne i}^K {\hat \beta _{{\rm{B - M}}}^{(k)}} }}{{({N_{{\rm{BS}}}} - 1)\hat \beta _{{\rm{B - M}}}^{(i)}}}{\rm{ + }}\frac{{{N_{{\rm{BS}}}}\sum\limits_{k \ne i}^K {\left( {\beta _{{\rm{B - M}}}^{(i)}{\rm{ - }}\hat \beta _{{\rm{B - M}}}^{(i)}} \right)\hat \beta _{{\rm{B - M}}}^{(k)}} }}{{({N_{{\rm{BS}}}} - 1)({N_{{\rm{BS}}}} - 2){{\left( {\hat \beta _{{\rm{B - M}}}^{(i)}} \right)}^2}}}.
\end{split}
\end{equation}
\end{table*}
Then, we define $\tilde \xi _{{\rm{B - M}}}^{(i)} \buildrel \Delta \over = \frac{{{{\left( {{\bf{\xi }}_{{\rm{B - M}}}^{(i)}} \right)}^H}{\hat { \bf g}}_{{\rm{B - M}}}^{(i)}}}{{\left\| {{\hat { \bf g}}_{{\rm{B - M}}}^{(i)}} \right\|}} \sim \mathcal{CN}\left( {0,\beta _{{\rm{B - M}}}^{(i)} - \hat \beta _{{\rm{B - M}}}^{(i)}} \right)$ and ${\hat {\tilde g}}_{{\rm{B - M}}}^{(k)} \buildrel \Delta \over = \frac{{{{\left( {{\hat { \bf g}}_{{\rm{B - M}}}^{(i)}} \right)}^H}{\hat { \bf g}}_{{\rm{B - M}}}^{(k)}}}{{\left\| {{\hat { \bf g}}_{{\rm{B - M}}}^{(i)}} \right\|}} \sim \mathcal{CN}\left( {0,\hat \beta _{{\rm{B - M}}}^{(k)}} \right)$. Conditioned on ${{\hat { \bf g}}_{{\rm{B - M}}}^{(i)}}$, $\tilde \xi _{{\rm{B - M}}}^{(i)}$ and ${\hat {\tilde g}}_{{\rm{B - M}}}^{(k)}$ are Gaussian distributed and independent of ${{\hat { \bf g}}_{{\rm{B - M}}}^{(i)}}$. Hence, we have (\ref{sec3:hat_mean_inv_PNI_MUE}) and (\ref{sec3:hat_mean_inv_IMI_MUE}). Also, law of large numbers~\cite{Cramer1970} leads to
\begin{equation}\label{sec3:hat_SCSI_rate_ext4}
    \begin{split}
        {\rm {E}}\left[ {{\rm{CT}}{{\rm{I}}_i}} \right] = \sum\limits_{n = 1}^S {{{\left( {\alpha _{{\rm{SC}}}^{(n)}} \right)}^2}{N_{{\rm{SC}}}}\Phi _{{\rm{S - S}}}^{(n)}\beta _{{\rm{S - M}}}^{(n,i)}}.
    \end{split}
\end{equation}
By substituting (\ref{sec3:hat_mean_inv_PNI_MUE}), (\ref{sec3:hat_mean_inv_IMI_MUE}) and (\ref{sec3:hat_SCSI_rate_ext4}) into (\ref{sec3:mean_inverse_MUE}), (\ref{sec3:hat_SCSI_MUE_rate2}) leads to
\begin{equation}\label{sec3:hat_mean_inv_MUE}
\begin{split}
&R_{{\rm{0,M,M}}}^{(i)} = \\
&{\log _2}\left( {1 + \frac{{\alpha _{{\rm{BS}}}^2({N_{{\rm{BS}}}} - 1)({N_{{\rm{BS}}}} - 2){{\left( {\hat \beta _{{\rm{B - M}}}^{(i)}} \right)}^2}}}{{\alpha _{{\rm{BS}}}^2 \hat \chi _{{\rm{BS}}}^{(i)} + \sum\limits_{n = 1}^S {{{\left( {\alpha _{{\rm{SC}}}^{(n)}} \right)}^2}{N_{{\rm{SC}}}}\Phi _{{\rm{S - S}}}^{(n)}\beta _{{\rm{S - M}}}^{(n,i)}}  + \sigma _0^2}}} \right)
\end{split}
\end{equation}
where $\hat \chi _{{\rm{BS}}}^{(i)} = {{\Phi }_{{\rm{B - M}}}}\left( {{N_{{\rm{BS}}}}\beta _{{\rm{B - M}}}^{(i)} - 2\hat \beta _{{\rm{B - M}}}^{(i)}} \right) - \left( {{N_{{\rm{BS}}}} - 4} \right){\left( {\hat \beta _{{\rm{B - M}}}^{(i)}} \right)^2} - 2\hat \beta _{{\rm{B - M}}}^{(i)}\beta _{{\rm{B - M}}}^{(i)}$.

For the derivation of the closed-form expression for the SUE achievable rate in (\ref{sec3:hat_SCSI_SUE_rate2_CF}), the expectation of the SINR's reciprocal can be written as (\ref{sec3:hat_mean_inverse_SUE}).
\begin{table*}
\begin{equation}\label{sec3:hat_mean_inverse_SUE}
\begin{split}
&{\rm {E}}\left[ {\frac{{{\rm{EE}}{{\rm{I}}_{m,j}} + {\rm{CT}}{{\rm{I}}_{m,j}} + {\rm{IS}}{{\rm{I}}_{m,j}} + {\rm{SS}}{{\rm{I}}_{m,j}} + \sigma _0^2}}{{{{\left( {\alpha _{{\rm{SC}}}^{(m)}} \right)}^2}{{\left\| {{\hat { \bf g}}_{{\rm{S - S}}}^{(m,m,j)}} \right\|}^4}}}} \right] = {\rm {E}}\left[ {\frac{{{\rm{EE}}{{\rm{I}}_{m,j}}}}{{{{\left( {\alpha _{{\rm{SC}}}^{(m)}} \right)}^2}{{\left\| {{\hat { \bf g}}_{{\rm{S - S}}}^{(m,m,j)}} \right\|}^4}}}} \right]\\
 &~~+ {\rm {E}}\left[ {\frac{{{\rm{IS}}{{\rm{I}}_{m,j}}}}{{{{\left( {\alpha _{{\rm{SC}}}^{(m)}} \right)}^2}{{\left\| {{\hat { \bf g}}_{{\rm{S - S}}}^{(m,m,j)}} \right\|}^4}}}} \right]+ {\rm {E}}\left[ {\frac{1}{{{{\left\| {{\hat { \bf g}}_{{\rm{S - S}}}^{(m,m,j)}} \right\|}^4}}}} \right]\frac{{{\rm {E}}\left[ {{\rm{CT}}{{\rm{I}}_{m,j}}} \right] + {\rm {E}}\left[ {{\rm{SS}}{{\rm{I}}_{m,j}}} \right] + \sigma _0^2}}{{{{\left( {\alpha _{{\rm{SC}}}^{(m)}} \right)}^2}}}
\end{split}
\end{equation}
\begin{equation}\label{sec3:mean_inv_24_SUE}
\begin{split}
&{\rm {E}}\left[ {\frac{1}{{{{\left\| {\hat{\bf{g}}_{{\rm{S - S}}}^{(m,m,j)}} \right\|}^2}}}} \right] = \frac{1}{{\left( {{N_{{\rm{SC}}}} - 1} \right)\hat\beta _{{\rm{S - S}}}^{(m,m,j)}}}, {\rm {E}}\left[ {\frac{1}{{{{\left\| {\hat{\bf{g}}_{{\rm{S - S}}}^{(m,m,j)}} \right\|}^4}}}} \right] = \frac{1}{{\left( {{N_{{\rm{SC}}}} - 1} \right)\left( {{N_{{\rm{SC}}}} - 2} \right){{\left( {\hat\beta _{{\rm{S - S}}}^{(m,m,j)}} \right)}^2}}}.
\end{split}
\end{equation}
\begin{equation}\label{sec3:hat_mean_inv_PNI_SUE}
\begin{split}
    {\rm {E}}\left[ {\frac{{{\rm{EE}}{{\rm{I}}_{m,j}}}}{{{{\left( {\alpha _{{\rm{SC}}}^{(m)}} \right)}^2}{{\left\| {{\hat { \bf g}}_{{\rm{S - S}}}^{(m,m,j)}} \right\|}^4}}}} \right] = {\rm {E}}\left[ {\frac{1}{{{{\left\| {{\hat { \bf g}}_{{\rm{S - S}}}^{(m,m,j)}} \right\|}^2}}}} \right]{\rm {E}}\left[ {{{\left| {\tilde \xi _{{\rm{S - S}}}^{(m,m,j)}} \right|}^2}} \right] = \frac{{\beta _{{\rm{S - S}}}^{(m,m,j)} - \hat \beta _{{\rm{S - S}}}^{(m,m,j)}}}{{\left( {{N_{{\rm{SC}}}} - 1} \right)\hat \beta _{{\rm{S - S}}}^{(m,m,j)}}}
\end{split}
\end{equation}
\begin{equation}\label{sec3:hat_mean_inv_ISI_SUE}
\begin{split}
    &{\rm {E}}\left[ {\frac{{{\rm{IS}}{{\rm{I}}_{m,j}}}}{{{{\left( {\alpha _{{\rm{SC}}}^{(m)}} \right)}^2}{{\left\| {{\hat { \bf g}}_{{\rm{S - S}}}^{(m,m,j)}} \right\|}^4}}}} \right] = {\rm {E}}\left[ {\frac{1}{{{{\left\| {{\hat { \bf g}}_{{\rm{S - S}}}^{(m,m,j)}} \right\|}^4}}}} \right]\sum\limits_{{l_1} \ne j}^L {{\rm {E}}\left[ {{{\left| {{{\left( {{\bf{\xi }}_{{\rm{S - S}}}^{(m,m,j)}} \right)}^T}{{\left( {{\hat { \bf g}}_{{\rm{S - S}}}^{(m,m,{l_1})}} \right)}^*}} \right|}^2}} \right]}\\
    &+{\rm {E}}\left[ {\frac{1}{{{{\left\| {{\hat { \bf g}}_{{\rm{S - S}}}^{(m,m,j)}} \right\|}^2}}}} \right]\sum\limits_{{l_1} \ne j}^L {{\rm {E}}\left[ {{{\left| {{\hat {\tilde g}}_{{\rm{S - S}}}^{(m,m,{l_1})}} \right|}^2}} \right]}= \frac{{\sum\limits_{{l_1} \ne j}^L {\hat \beta _{{\rm{S - S}}}^{(m,m,{l_1})}} }}{{\left( {{N_{{\rm{SC}}}} - 1} \right)\hat \beta _{{\rm{S - S}}}^{(m,m,j)}}} + \frac{{{N_{{\rm{SC}}}}\sum\limits_{{l_1} \ne j}^L {\left( {\beta _{{\rm{S - S}}}^{(m,m,j)} - \hat \beta _{{\rm{S - S}}}^{(m,m,j)}} \right)\hat \beta _{{\rm{S - S}}}^{(m,m,{l_1})}} }}{{\left( {{N_{{\rm{SC}}}} - 1} \right)\left( {{N_{{\rm{SC}}}} - 2} \right){{\left( {\hat \beta _{{\rm{S - S}}}^{(m,m,j)}} \right)}^2}}}
\end{split}
\end{equation}
\begin{equation}\label{sec3:hat_SCSI_rate_ext7}
    \begin{split}
        {\rm {E}}\left[ {{\rm{CT}}{{\rm{I}}_{m,j}}} \right] = \alpha _{{\rm{BS}}}^2{N_{{\rm{BS}}}}{{\Phi }_{{\rm{B - M}}}}\beta _{{\rm{B - S}}}^{(m,j)},{\rm {E}}\left[ {{\rm{SS}}{{\rm{I}}_{m,j}}} \right] = \sum\limits_{n \ne m}^S {{{\left( {\alpha _{{\rm{SC}}}^{(n)}} \right)}^2}{N_{{\rm{SC}}}}\Phi _{{\rm{S - S}}}^{(n)}\beta _{{\rm{S - S}}}^{(n,m,j)}} + \sum\limits_{\scriptstyle n \ne m\hfill\atop
\scriptstyle n \in {{\mathcal A}_r}\hfill} {{\left( {\alpha _{{\rm{SC}}}^{(n)}} \right)}^2}{N_{{\rm{SC}}}^2}\beta _{{\rm{S - S}}}^{(n,m,j)}\hat\beta _{{\rm{S - S}}}^{(n,m,j)}
    \end{split}
\end{equation}
\begin{equation}\label{sec3:hat_mean_inv_SUE}
\begin{split}
R_{{\rm{0,S,M}}}^{(m,j)} = {\log _2}\left( {1 + \frac{{{{\left( {\alpha _{{\rm{SC}}}^{(m)}} \right)}^2}\left( {{N_{{\rm{SC}}}} - 1} \right)\left( {{N_{{\rm{SC}}}} - 2} \right){{\left( {\beta _{{\rm{S - S}}}^{(m,m,j)}} \right)}^2}}}{{{{\left( {\alpha _{{\rm{SC}}}^{(m)}} \right)}^2}\hat \chi _{{\rm{SC}}}^{\left( {m,j} \right)} + \alpha _{{\rm{BS}}}^2\hat \chi _{{\rm{BS}}}^{\left( {m,j} \right)} + \sum\limits_{n \ne m}^S {{{\left( {\alpha _{{\rm{SC}}}^{(n)}} \right)}^2}{\hat \chi _{{\rm{SC}},1}^{\left( {n,m,j} \right)}}} + \sum\limits_{\scriptstyle n \ne m\hfill\atop
\scriptstyle n \in {{\mathcal A}_r}\hfill} {{\left( {\alpha _{{\rm{SC}}}^{(n)}} \right)}^2}{\hat \chi _{{\rm{SC}},2}^{\left( {n,m,j} \right)}}  + \sigma _0^2}}} \right)
\end{split}
\end{equation}
\begin{equation}\label{sec3:hat_mean_inverse_MUE2}
    \begin{split}
    &{\rm {E}}\left[ {\frac{{{\rm{EE}}{{\rm{I}}_i} + {\rm{IM}}{{\rm{I}}_i} + {\rm{CT}}{{\rm{I}}_i} + \sigma _0^2}}{{\alpha _{{\rm{BS}}}^2}}} \right] =~ \frac{{\rm {E}}\left[ {\rm{EE}}{{\rm{I}}_i}+{\rm{IM}}{{\rm{I}}_i} \right]+{\rm {E}}\left[ {{\rm{CT}}{{\rm{I}}_i}} \right] + \sigma _0^2}{{\alpha _{{\rm{BS}}}^2}}.
    \end{split}
\end{equation}
\begin{equation}\label{sec3:hat_mean_inv_PNI_MUE2}
\begin{split}
&{\rm {E}}\left[ {\rm{EE}}{{\rm{I}}_i}+{\rm{IM}}{{\rm{I}}_i} \right] = \alpha _{{\rm{BS}}}^2 {\rm {E}}\left[ {{\left( {{\bf{\xi}}_{{\rm{B - M}}}^{(i)}} \right)}^H}{\hat { \bar {\bf G}}}_{{\rm{B - M}}}{\hat { \bar {\bf G}}}_{{\rm{B - M}}}^H {{\bf{\xi}}_{{\rm{B - M}}}^{(i)}} \right]= \alpha _{{\rm{BS}}}^2 \left({{\beta}_{{\rm{B - M}}}^{(i)}}-{{\hat{\beta}}_{{\rm{B - M}}}^{(i)}}\right)\frac{{\Psi}_{\rm B-M}}{N_{\rm BS}-K-1}
\end{split}
\end{equation}
\begin{equation}\label{sec3:hat_mean_inv_IMI_MUE2}
\begin{split}
&{\rm {E}}\left[ {\rm{CT}}{{\rm{I}}_i} \right] = \sum\limits_{n = 1}^S {\left( {\alpha _{{\rm{SC}}}^{(n)}} \right)}^2 {{\beta}_{{\rm{S - S}}}^{(n,i)}}{\rm {Tr}}\left\{{\rm {E}}\left[ \left[\left({\hat {{\bf G}}}_{{\rm{S - S}}}^{(n,n)}\right)^H{\hat {{\bf G}}}_{{\rm{S - S}}}^{(n,n)}\right]^{-1} \right]\right\}= \sum\limits_{n = 1}^S {\left( {\alpha _{{\rm{SC}}}^{(n)}} \right)}^2 {{\beta}_{{\rm{S - S}}}^{(n,i)}}\frac{{\Psi}_{\rm S-S}^{(n)}}{N_{\rm SC}-L-1}
\end{split}
\end{equation}
\begin{equation}\label{sec3:hat_mean_inv_MUE2}
\begin{split}
R_{{\rm{0,M,Z}}}^{(i)} = {\log _2}\left( {1 + \frac{{\alpha _{{\rm{BS}}}^2}}{{\alpha _{{\rm{BS}}}^2\frac{\left({{\beta}_{{\rm{B - M}}}^{(i)}}-{{\hat{\beta}}_{{\rm{B - M}}}^{(i)}}\right){\Psi}_{\rm B-M}}{N_{\rm BS}-K-1}+\sum\limits_{n = 1}^S {\left( {\alpha _{{\rm{SC}}}^{(n)}} \right)}^2 {{\beta}_{{\rm{S - S}}}^{(n,i)}}\frac{{\Psi}_{\rm S-S}^{(n)}}{N_{\rm SC}-L-1}+ \sigma _0^2}}} \right).
\end{split}
\end{equation}
\end{table*}
where $\frac{{{\rm {E}}\left[ {{\rm{CT}}{{\rm{I}}_{m,j}}} \right] + {\rm {E}}\left[ {{\rm{SS}}{{\rm{I}}_{m,j}}} \right] + \sigma _0^2}}{{{{\left( {\alpha _{{\rm{SC}}}^{(m)}} \right)}^2}}}$ is independent of ${{\hat { \bf g}}_{{\rm{S - S}}}^{(m,m,j)}}$. According to the \cite[Lemma~2.9]{Tulino2004} given in (\ref{sec3:Lemma2_9}), we have (\ref{sec3:mean_inv_24_SUE}). Likewise, we define $\tilde \xi _{{\rm{S - S}}}^{(m,m,j)} \buildrel \Delta \over = \frac{{{{\left( {{\bf{\xi }}_{{\rm{S - S}}}^{(m,m,j)}} \right)}^H}{\hat { \bf g}}_{{\rm{S - S}}}^{(m,m,j)}}}{{\left\| {{\hat { \bf g}}_{{\rm{S - S}}}^{(m,m,j)}} \right\|}} \sim CN\left( {0,\beta _{{\rm{S - S}}}^{(m,m,j)} - \hat \beta _{{\rm{S - S}}}^{(m,m,j)}} \right)$ and ${\hat {\tilde g}}_{{\rm{S - S}}}^{(m,m,{l_1})} \buildrel \Delta \over = \frac{{{{\left( {{\hat { \bf g}}_{{\rm{S - S}}}^{(m,m,j)}} \right)}^H}{\hat { \bf g}}_{{\rm{S - S}}}^{(m,m,{l_1})}}}{{\left\| {{\hat { \bf g}}_{{\rm{S - S}}}^{(m,m,j)}} \right\|}} \sim CN\left( {0,\hat \beta _{{\rm{S - S}}}^{(m,m,{l_1})}} \right)$, both of which do not depend on ${{\hat { \bf g}}_{{\rm{S - S}}}^{(m,m,j)}}$ conditioned on it, then we obtain (\ref{sec3:hat_mean_inv_PNI_SUE}) and (\ref{sec3:hat_mean_inv_ISI_SUE}). Similarly, the law of large numbers~\cite{Cramer1970} leads to (\ref{sec3:hat_SCSI_rate_ext7}), where $\hat \beta _{{\rm{S - S}}}^{(n,m,j)} \buildrel \Delta \over = \frac{{\tau {p_{\tau}}{{\left( {\beta _{{\rm{S - S}}}^{(n,m,j)}} \right)}^2}}}{{\tau {p_{\tau}}\sum\limits_{l \in {{\mathcal A}_r}} {\beta _{{\rm{S - S}}}^{(n,l,j)}}  + \sigma _0^2}}$. Then, substituting (\ref{sec3:hat_mean_inv_PNI_SUE}), (\ref{sec3:hat_mean_inv_ISI_SUE}), and (\ref{sec3:hat_SCSI_rate_ext7}) into (\ref{sec3:mean_inv_24_SUE}), (\ref{sec3:hat_SCSI_SUE_rate2}) leads to (\ref{sec3:hat_mean_inv_SUE}), where $\hat \chi _{{\rm{SC}}}^{\left( {m,j} \right)} = \Phi _{{\rm{S - S}}}^{(m)}\left( {{N_{{\rm{SC}}}}\beta _{{\rm{S - S}}}^{(m,m,j)} - 2\hat \beta _{{\rm{S - S}}}^{(m,m,j)}} \right) - \left( {{N_{{\rm{SC}}}} - 4} \right){\left( {\hat \beta _{{\rm{S - S}}}^{(m,m,j)}} \right)^2} - 2\beta _{{\rm{S - S}}}^{(m,m,j)}\hat \beta _{{\rm{S - S}}}^{(m,m,j)}$, $\hat \chi _{{\rm{BS}}}^{\left( {m,j} \right)} = {N_{{\rm{BS}}}}{\Phi _{{\rm{B - M}}}}\beta _{{\rm{B - S}}}^{(m,j)}$, $\hat \chi _{{\rm{SC}},1}^{\left( {n,m,j} \right)} = {N_{{\rm{SC}}}}\Phi _{{\rm{S - S}}}^{(n)}\beta _{{\rm{S - S}}}^{(n,m,j)}$, and $\hat \chi _{{\rm{SC}},2}^{\left( {n,m,j} \right)} = N_{{\rm{SC}}}^2\beta _{{\rm{S - S}}}^{(n,m,j)}\hat \beta _{{\rm{S - S}}}^{(n,m,j)}$.

Finally, by substituting (\ref{sec3:hat_alpha1}) into (\ref{sec3:hat_mean_inv_MUE}) and (\ref{sec3:hat_mean_inv_SUE}), (\ref{sec3:hat_SCSI_MUE_rate2_CF}) and (\ref{sec3:hat_SCSI_SUE_rate2_CF}) are obtained and thus Theorem 1 is demonstrated.

\section{Proof of Theorem 2}\label{sec3:Appendix_D}
Similar to the proof of Theorem 1 in Appendix~\ref{sec3:Appendix_B}, to derive the closed-form expression of the ZFT based MUE achievable rate in (\ref{sec3:hat_SCSI_MUE_rate2_CF2}) based on Jensen's inequality, we start from the expectation of the SINR's reciprocal, given by (\ref{sec3:hat_mean_inverse_MUE2}).

Then, the property of the inverted Wishart Distribution in (\ref{sec3:inv_Wishart}), the law of large numbers~\cite{Cramer1970} and ${\rm {Tr}}\left\{ {{\bf{AB}}} \right\} = {\rm {Tr}}\left\{ {{\bf{BA}}} \right\}$ offer us (\ref{sec3:hat_mean_inv_PNI_MUE2}) and (\ref{sec3:hat_mean_inv_IMI_MUE2}), where ${{\Psi }_{{\rm{B - M}}}} = \sum\limits_{i = 1}^K {\frac{1}{{\hat \beta _{{\rm{B - M}}}^{(i)}}}}$, and $\Psi _{{\rm{S - S}}}^{(n)} = \sum\limits_{l = 1}^L {\frac{1}{{\hat \beta _{{\rm{S - S}}}^{(n,n,l)}}}}$ with $n \in\left\{1,\dots,S\right\}$.

By substituting (\ref{sec3:hat_mean_inv_PNI_MUE2}) and (\ref{sec3:hat_mean_inv_IMI_MUE2}) into (\ref{sec3:hat_mean_inverse_MUE2}), (\ref{sec3:hat_SCSI_MUE_rate2}) leads to (\ref{sec3:hat_mean_inv_MUE2}).

For the derivation of the closed-form expression for the SUE achievable rate in (\ref{sec3:hat_SCSI_SUE_rate2_CF2}), the expectation of the SINR's reciprocal can be written as
\begin{equation}\label{sec3:hat_mean_inverse_SUE2}
\begin{split}
&{\rm {E}}\left[ {\frac{{{\rm{EE}}{{\rm{I}}_{m,j}} + {\rm{CT}}{{\rm{I}}_{m,j}} + {\rm{IS}}{{\rm{I}}_{m,j}} + {\rm{SS}}{{\rm{I}}_{m,j}} + \sigma _0^2}}{{{{\left( {\alpha _{{\rm{SC}}}^{(m)}} \right)}^2}}}} \right]\\
=~&\frac{{\rm {E}}\left[ {\rm{EE}}{{\rm{I}}_{m,j}} + {\rm{IS}}{{\rm{I}}_{m,j}} \right] + {\rm {E}}\left[ {\rm{CT}}{{\rm{I}}_{m,j}}\right] + {\rm {E}}\left[ {\rm{SS}}{{\rm{I}}_{m,j}} \right] + \sigma _0^2}{{\left( {\alpha _{{\rm{SC}}}^{(m)}} \right)}^2}.
\end{split}
\end{equation}

Similarly, according to the property of the inverted Wishart Distribution in (\ref{sec3:inv_Wishart}), we have (\ref{sec3:hat_mean_inv_PNI_SUE2}), (\ref{sec3:hat_mean_inv_MSI_SUE2}) and (\ref{sec3:hat_mean_inv_SSI_SUE2}).
\begin{table*}
\begin{equation}\label{sec3:hat_mean_inv_PNI_SUE2}
\begin{split}
    {\rm {E}}\left[{{\rm{EE}}{{\rm{I}}_{m,j}}}+{{\rm{IS}}{\rm{I}}_{m,j}}\right] & = {\rm {E}}\left[ {\left( {\alpha _{{\rm{SC}}}^{(m)}} \right)^2}{\left| {{{\left( {{\bf{\xi }}_{{\rm{S - S}}}^{(m,m,j)}} \right)}^H}{\hat {\bar{\bf g}}}_{{\rm{S - S}}}^{(m,m,j)}} \right|^2} + \sum\limits_{{l_1} \ne j}^L {{{\left( {\alpha _{{\rm{SC}}}^{(m)}} \right)}^2}{{\left| {{{\left( {{\bf{\xi}}_{{\rm{S - S}}}^{(m,m,j)}} \right)}^H}{{\hat {\bar{\bf g}}}_{{\rm{S - S}}}^{(m,m,{l_1})}}} \right|}^2}}\right]\\
    &= \frac{{\left( {\alpha _{{\rm{SC}}}^{(m)}} \right)^2} \left({{\beta}_{{\rm{S - S}}}^{(m,m,j)}}-{{\hat{\beta}}_{{\rm{S - S}}}^{(m,m,j)}}\right){\Psi}_{\rm S-S}^{(m)}}{N_{\rm SC}-L-1}
\end{split}
\end{equation}
\begin{equation}\label{sec3:hat_mean_inv_MSI_SUE2}
\begin{split}
    {\rm {E}}\left[ {\rm{CT}}{{\rm{I}}_{m,j}} \right] = {\rm {E}}\left[ \sum\limits_{i = 1}^K {\alpha _{{\rm{BS}}}^2{{\left| {{{\left( {{\bf{g}}_{{\rm{B - S}}}^{(m,j)}} \right)}^H}{{\hat {\bar{\bf g}}}_{{\rm{B - M}}}^{(i)}}} \right|}^2}} \right] = \frac{\alpha _{{\rm{BS}}}^2 {{\beta}_{{\rm{B - S}}}^{(m,j)}} \Psi_{\rm B-M}}{N_{\rm BS}-K-1}
\end{split}
\end{equation}
\begin{equation}\label{sec3:hat_mean_inv_SSI_SUE2}
\begin{split}
    {\rm {E}}\left[ {\rm{SS}}{{\rm{I}}_{m,j}} \right] &= \sum\limits_{n \ne m}^S {{\left( {\alpha _{{\rm{SC}}}^{(n)}} \right)}^2} {\rm {E}}\left[{\sum\limits_{{l_2} = 1}^L {{{\left| {{{\left( {{\bf{g}}_{{\rm{S - S}}}^{(n,m,j)}} \right)}^H}{{\hat {\bar{\bf g}}}_{{\rm{S - S}}}^{(n,n,{l_2})}}} \right|}^2}} } \right] = \frac{ \sum\limits_{n \ne m}^S{{\left( {\alpha _{{\rm{SC}}}^{(n)}} \right)}^2} {{\beta}_{{\rm{S - S}}}^{(n,m,j)}}\Psi_{\rm S-S}^{(n)}}{N_{\rm SC}-L-1} \\
     &+ \sum\limits_{\scriptstyle n \ne m\hfill\atop
\scriptstyle n \in {{\mathcal A}_r}\hfill}{{\left( {\alpha _{{\rm{SC}}}^{(n)}} \right)}^2}\left[ {{{\left( {\frac{{\beta _{{\rm{S - S}}}^{(n,m,j)}}}{{\beta _{{\rm{S - S}}}^{(n,n,j)}}}} \right)}^2} - \frac{{\beta _{{\rm{S - S}}}^{(n,m,j)}}}{{\left( {{N_{{\rm{SC}}}} - L - 1} \right)\hat \beta _{{\rm{S - S}}}^{(n,n,j)}}}} \right].
\end{split}
\end{equation}
\begin{equation}\label{sec3:hat_mean_inv_SUE2}
\begin{split}
R_{{\rm{0,S,Z}}}^{(m,j)} = {\log _2}\left( 1 + \frac{{\left( {\alpha _{{\rm{SC}}}^{(m)}} \right)}^2}{{\left( {\alpha _{{\rm{SC}}}^{(m)}} \right)^2}\hat\eta _{{\rm{SC}}}^{\left( {m,j} \right)}+ {\alpha _{{\rm{BS}}}^2\hat \eta _{{\rm{BS}}}^{\left( {m,j} \right)}} + {\sum\limits_{n \ne m}^S {{{\left( {\alpha _{{\rm{SC}}}^{(n)}} \right)}^2}\hat \eta _{{\rm{SC,1}}}^{\left( {n,m,j} \right)}} } + \sum\limits_{\scriptstyle n \ne m\hfill\atop
\scriptstyle n \in {{\mathcal A}_r}\hfill}{{\left( {\alpha _{{\rm{SC}}}^{(n)}} \right)}^2} \hat\eta _{{\rm{SC}},2}^{\left( {n,m,j} \right)}+ \sigma _0^2} \right)
\end{split}
\end{equation}
\end{table*}

Then, substituting (\ref{sec3:hat_mean_inv_PNI_SUE2}), (\ref{sec3:hat_mean_inv_MSI_SUE2}) and (\ref{sec3:hat_mean_inv_SSI_SUE2}) into (\ref{sec3:hat_mean_inverse_SUE2}), (\ref{sec3:hat_SCSI_SUE_rate2}) leads to (\ref{sec3:hat_mean_inv_SUE2}), where $\hat\eta _{{\rm{SC}}}^{\left( {m,j} \right)} = \frac{{\left( {\beta _{{\rm{S - S}}}^{(n,m,j)} - \hat \beta _{{\rm{S - S}}}^{(m,m,j)}} \right)\Psi _{{\rm{S - S}}}^{(m)}}}{{{N_{{\rm{SC}}}} - L - 1}}$, $\hat \eta _{{\rm{BS}}}^{\left( {m,j} \right)} = \frac{{\beta _{{\rm{B - S}}}^{(m,j)}{\Psi _{{\rm{B - M}}}}}}{{{N_{{\rm{BS}}}} - K - 1}}$, $\hat \eta _{{\rm{SC,1}}}^{\left( {n,m,j} \right)} = \frac{{\beta _{{\rm{S - S}}}^{(n,m,j)}\Psi _{{\rm{S - S}}}^{(n)}}}{{{N_{{\rm{SC}}}} - L - 1}}$, and $\hat\eta _{{\rm{SC}},2}^{\left( {n,m,j} \right)} = \left[ {{{\left( {\frac{{\beta _{{\rm{S - S}}}^{(n,m,j)}}}{{\beta _{{\rm{S - S}}}^{(n,n,j)}}}} \right)}^2} - \frac{{\beta _{{\rm{S - S}}}^{(n,m,j)}}}{{\left( {{N_{{\rm{SC}}}} - L - 1} \right)\hat \beta _{{\rm{S - S}}}^{(n,n,j)}}}} \right]$.
Finally, by substituting (\ref{sec3:hat_alpha2}) into (\ref{sec3:hat_mean_inv_MUE2}) and (\ref{sec3:hat_mean_inv_SUE2}), (\ref{sec3:hat_SCSI_MUE_rate2_CF2}) and (\ref{sec3:hat_SCSI_SUE_rate2_CF2}) are obtained and thus Theorem 2 is demonstrated.

%
%
\end{document}